\documentclass[a4paper,11pt]{article}

\pdfoutput=1

\usepackage{amsmath,amssymb,mathtools}
\usepackage{color}
\usepackage{graphicx}
\usepackage{subfigure}
\usepackage{cite}
\usepackage{hyperref}
\usepackage{multirow,makecell}
\usepackage{textcomp}
\usepackage{wasysym}

\usepackage[text={17.1cm,24.6cm},centering]{geometry} 

\numberwithin{equation}{section}

\def \be {\begin{equation}}
\def \ee {\end{equation}}
\def \ba {\begin{array}}
\def \ea {\end{array}}
\def \bea{\begin{eqnarray}}
\def \eea{\end{eqnarray}}
\def \nn {\nonumber}

\def \a {\alpha}
\def \b {\beta}

\def \G {\Gamma}
\def \d {\delta}
\def \D {\Delta}
\def \e {\epsilon}

\def \m {\mu}

\def \s {\sigma}

\def \r {\rho}

\def \O {\Omega}
\def \th {\theta}

\def \mA {\mathcal A}
\def \mB {\mathcal B}
\def \mC {\mathcal C}
\def \mD {\mathcal D}
\def \mE {\mathcal E}
\def \mF {\mathcal F}
\def \mG {\mathcal G}
\def \mH {\mathcal H}
\def \mI {\mathcal I}
\def \mJ {\mathcal J}

\def \mO {\mathcal O}

\def \mR {\mathcal R}
\def \mS {\mathcal S}
\def \mT {\mathcal T}

\def \mX {\mathcal X}
\def \mY {\mathcal Y}

\def \p {\partial}
\def \es {\emptyset}
\def \f {\frac}

\def \lt {\left}
\def \rt {\right}

\def \ra {\rightarrow}
\def \lra {\leftrightarrow}
\def \sr {\sqrt}
\def \td {\tilde}

\def \pp {\propto}
\def \inf {\infty}

\def \lag {\langle}
\def \rag {\rangle}

\def \ep {\mathrm{e}}
\def \ii {\mathrm{i}}

\def \Re {{\textrm{Re}}}

\def \tr {\textrm{tr}}

\def \and {{~\textrm{and}~}}

\def \CFT {{\textrm{CFT}}}

\def \GGE {{\textrm{GGE}}}

\def \me {{\textrm{me}}}

\def \Z {{\textrm{Z}}}

\def \cL {{\textrm{L}}}
\def \NL {{\textrm{NL}}}
\def \NNL {{\textrm{NNL}}}

\begin{document}

\title{\textbf{Dissimilarities of reduced density matrices and eigenstate thermalization hypothesis}}
\author{
Song He$^{1,2}$\footnote{hesong17@gmail.com}~,
Feng-Li Lin$^{3}$\footnote{fengli.lin@gmail.com}~
and
Jia-ju Zhang$^{4,5}$\footnote{jiaju.zhang@mib.infn.it}
}
\date{}

\maketitle
\vspace{-10mm}
\begin{center}
{\it
$^{1}$Max Planck Institute for Gravitational Physics (Albert Einstein Institute),\\
Am M\"uhlenberg 1, 14476 Golm, Germany\\\vspace{2mm}
$^{2}$CAS Key Laboratory of Theoretical Physics, Institute of Theoretical Physics, Chinese Academy of Sciences, 55 Zhong Guan Cun East Road, Beijing 100190, China\\\vspace{2mm}
$^{3}$Department of Physics, National Taiwan Normal University,\\
No.\ 88, Sec.\ 4, Ting-Chou Road, Taipei 11677, Taiwan\\\vspace{2mm}
$^{4}$Dipartimento di Fisica, Universit\`a degli Studi di Milano-Bicocca,\\Piazza della Scienza 3, I-20126 Milano, Italy\\\vspace{2mm}
$^{5}$INFN, Sezione di Milano-Bicocca, Piazza della Scienza 3, I-20126 Milano, Italy\\\vspace{2mm}
}
\vspace{10mm}
\end{center}

\begin{abstract}

  We calculate various quantities that characterize the dissimilarity of reduced density matrices for a short interval of  length $\ell$ in a two-dimensional (2D) large central charge conformal field theory (CFT). These quantities include the R\'enyi entropy, entanglement entropy, relative entropy, Jensen-Shannon divergence, as well as the Schatten 2-norm and 4-norm. We adopt the method of operator product expansion of twist operators, and calculate the short interval expansion of these quantities up to order of $\ell^9$ for the contributions from the vacuum conformal family. The formal forms of these dissimilarity measures and the derived Fisher information metric from contributions of general operators are also given. As an application of the results, we use these dissimilarity measures to compare the excited and thermal states, and examine the eigenstate thermalization hypothesis (ETH) by showing how they behave in high temperature limit. This would help to understand how ETH in 2D CFT can be defined more precisely. We discuss the possibility that all the dissimilarity measures considered here vanish when comparing the reduced density matrices of an excited state and a generalized Gibbs ensemble thermal state. We also discuss ETH for a microcanonical ensemble thermal state in a 2D large central charge CFT, and find that it is approximately satisfied for a small subsystem and violated for a large subsystem.

\end{abstract}

\baselineskip 18pt
\thispagestyle{empty}
\newpage

\tableofcontents


\section{Introduction}

Motivated by the eigenstate thermalization hypothesis (ETH) \cite{Deutsch:1991,Srednicki:1994} or its generalization, the subsystem ETH \cite{Lashkari:2016vgj,Dymarsky:2016aqv}, it is important to characterize quantitatively the difference between the excited state and the thermal state. One such characterization is to quantify the difference between reduced density matrices over a local regions of these two states. This is also an interesting question by itself in quantum information theory. For two-dimensional (2D) conformal field theory (CFT), many other quantities of examining ETH have been adopted, such as correlation functions \cite{Fitzpatrick:2014vua,Fitzpatrick:2015zha}, entanglement entropy, R\'enyi entropy, relative entropy \cite{Asplund:2014coa,Caputa:2014eta,Lashkari:2016vgj,Lin:2016dxa,Dymarsky:2016aqv,He:2017vyf}, trace square \cite{Basu:2017kzo}, etc. Due to the infinite number of degrees of freedom in CFT, not every quantity is good for the use of examining the ETH \cite{Lashkari:2016vgj,Dymarsky:2016aqv}, unless its behaviors for both excited and thermal states are known precisely.

It was proposed in \cite{Calabrese:2004eu} to use correlation functions of twist operators to calculate the R\'enyi entropy in a 2D CFT, i.e., the partition function of the Riemann surface resulting from the replica trick.
When there is no compact form for these twist-operator correlation functions, one can use operator product expansion (OPE) of twist operators to calculate the short interval expansion of R\'enyi entropy \cite{Calabrese:2009ez,Headrick:2010zt,Calabrese:2010he,Chen:2013kpa,Chen:2016lbu}.  Following this method, in this paper we will calculate various quantities which are just the sums of some partition functions, and moreover can be used to characterize the dissimilarity of the reduced density matrices of thermal and excited states, and other states on various Riemann surfaces.

Our results can be used to examine ETH. The ETH and subsystem ETH are originally defined by comparing the highly excited state with the microcanonical ensemble thermal state \cite{Deutsch:1991,Srednicki:1994,Lashkari:2016vgj,Dymarsky:2016aqv}. Motivated by \cite{Iyoda:2016,Tasaki:2016}, as well as \cite{Fitzpatrick:2014vua,Fitzpatrick:2015zha,Asplund:2014coa,Caputa:2014eta}, we compare in \cite{He:2017vyf} the excited state with the canonical ensemble thermal state, and adopt the so-called weak ETH \cite{Iyoda:2016,Tasaki:2016}.
In \cite{He:2017vyf} the short-interval $\ell$ expansions of the entanglement entropies for the excited state and canonical ensemble thermal state are calculated to order $\ell^8$, and it was found that their difference, which is just the relative entropy, is only suppressed by the powers of large central charge $c$, instead of exponential suppression.
In this paper we show that there are similar behaviors for the Jensen-Shannon divergence and Schatten 2-norm.
For the more refined consideration, one should compare the excited state with the generalized Gibbs ensemble (GGE) thermal state \cite{Rigol:2006,Mandal:2015jla,Cardy:2015xaa,Mandal:2015kxi,Vidmar:2016,deBoer:2016bov}.
We will discuss the possibility that all the dissimilarities considered in this paper vanish when comparing the reduced density matrices of an excited state and a suitably defined GGE thermal state. As a by-product, we also check ETH for the microcanonical ensemble thermal state with the dissimilarity measures of comparing with the energy eigenstate.

The rest of this paper is arranged as follows.
In section~\ref{secpre} we give prescriptions of the method and show how to get the partition functions from OPE of twist operators. Moreover,  in subsection~\ref{secren} we  apply the prescriptions to evaluate the R\'enyi and entanglement entropies.
In section~\ref{secmeasure} we calculate the various dissimilarity measures between reduced density matrices.
In section~\ref{secethce} we apply our results to examine ETH for the canonical ensemble thermal state.
In section~\ref{secethgge} we discuss the possible scenarios ETH for the GGE thermal state.
In section~\ref{secme} we discuss ETH for a microcanonical ensemble thermal state in a 2D large central charge CFT, and find that it is approximately satisfied for a small subsystem and violated for a large subsystem.
We conclude with discussion in section~\ref{seccon}.
In appendix~\ref{apprel} we calculate the relative entropy from modular Hamiltonian as a consistent check.
In appendix~\ref{secgen} we consider the contributions from general operators, and get the formal forms of the various dissimilarity measures and the Fisher information metric.
Some lengthy and not so enlightening results in section~\ref{secmeasure} are collected in appendix~\ref{appcl}.

\section{Prescriptions of the method}\label{secpre}

In this section we first give the useful basics of the vacuum conformal family in two-dimensional large central charge CFT and then show how we calculate the partition functions on various Riemann surfaces using OPE of the twist operators.

\subsection{CFT basics}

In this paper we only consider the contributions from the holomorphic sector of the vacuum conformal family in a two-dimensional large central charge CFT, and the generalization to antiholomorphic sector can be figured out easily.
We need the quasiprimary operators to level 9, i.e.,  $T$, $\mA$, $\mB$, $\mD$, $\mE$, $\mH$, $\mI$ and $\mJ$ as shown in table~\ref{vacconf}. The definitions, normalization factors, and conformal transformations of the quasiprimary operators up to level 8, as well as some useful structure constants, can be found in \cite{Chen:2013kpa,Chen:2013dxa,He:2017vyf,Chen:2016lbu}.

\begin{table}[htbp]
  \centering
\begin{tabular}{|c|c|c|c|c|c|c|}\hline
  level    & 0 & 2   & 4     & 6            & 8                   & 9     \\ \hline
  operator & 1 & $T$ & $\mA$ & $\mB$, $\mD$ & $\mE$, $\mH$, $\mI$ & $\mJ$ \\ \hline
\end{tabular}
  \caption{The holomorphic quasiprimary operators to level 9 in vacuum conformal family of a two-dimensional large central charge CFT.}\label{vacconf}
\end{table}

In this paper, we need the additional structure constants
\bea \label{z18}
&& C_{TT\mE} = \frac{20 c (105 c+11)}{63}, ~~ C_{TT\mH} = C_{TT\mI} = 0, \nn\\
&& C_{\mA\mA\mE} = \frac{8 c (5 c+22) (525 c+2419)}{315}, \nn\\
&& C_{\mA\mA\mH} = -\frac{8 c (5 c+22) (8400 c^2+44575 c-6961)}{125 (105 c+11)} ,\\
&& C_{\mA\mA\mI} = \frac{3 c (2 c-1) (3 c+46) (5 c+3) (5 c+22) (7 c+68)}{2(1050 c^2+3305 c-251)}.\nn
\eea
Furthermore, at level 9 we have the operator and its normalization
\bea \label{z13}
&& \hspace{-8mm} -\ii \mJ = (\p T(\p T \p T)) - \f65 (\p^2T(\p TT))
                           + \f{4}{15}(\p^3T(TT)) - \f{1}{10}(\p^4T\p T)
                           + \f{1}{100}(\p^5TT) - \f{1}{3150}\p^7T, \nn\\
&& \hspace{-8mm} \a_\mJ = \f{224c(2c-1)(5c+22)}{25},
\eea
with $(\mX\mY)$ denoting normal ordering of two operators $\mX$ and $\mY$.
Under a general conformal transformation $z \to f(z)$ it transforms as
\be \label{z14}
\mJ(z) = f'^9 \mJ(f) + \cdots + \frac{c (2 c-1) (5 c+22) (4 s^2 s'''+15 s'^3-18 s s' s'' )}{259200},
\ee
where $s$ denotes the Schwarzian derivative
\be
s(x) = \f{f'''(x)}{f'(x)} - \f32 \Big( \f{f''(x)}{f'(x)} \Big)^2,
\ee
and $\cdots$ represents the omitted terms that are proportional to $T$, $\mA$, $\mB$, $\mD$ and their derivatives.

\subsection{OPE of twist operators}

For one short interval $A=[0,\ell]$ on a Riemann surface $\mR$, replica trick leads to a CFT on an $n$-fold Riemann surface $\mR^n$. The partition function on $\mR^n$ can be written as a two-point function of twist operators $\mT$ and $\td\mT$ in an $n$-fold CFT on $\mR$ \cite{Calabrese:2004eu}
\be \label{z21}
\tr_A\r_A^n = \lag \mT(\ell)\td\mT(0)\rag_\mR, ~~ h_\mT=h_{\td\mT}=\f{c(n^2-1)}{24n},
\ee
and the $n$ folds of the CFT, which we call CFT$^n$, are independent except the connection by the twist operators. 
In this paper we only consider Riemann surface $\mR$ with translation symmetry, and so the one-point functions are all constants. Using OPE of twist operators \cite{Calabrese:2009ez,Headrick:2010zt,Calabrese:2010he,Chen:2013kpa,Chen:2016lbu}, we may get
\be \label{z15}
\lag \mT(\ell)\td \mT(0) \rag_\mR =\f{c_n}{\ell^{2h_\s}} \sum_K d_K \ell^{h_K} \lag\Phi_K(0)\rag_\mR,
\ee
and in the summation we only need to consider the quasiprimary operators $\Phi_K$ in CFT$^n$ that are the direct products of the quasiprimary operators in different replicas of the CFT. Only considering the contributions from the vacuum conformal family, we list the quasiprimary operators in CFT$^n$ to level 9 in table~\ref{vacconfn}. To level 8, the coefficients $d_K$ can be found in \cite{Chen:2013kpa,Chen:2013dxa}, and using the method in \cite{Calabrese:2010he} and (\ref{z13}), (\ref{z14}) we can easily get
\be \label{z16}
d_\mJ = 0.
\ee
Interestingly, there is no contribution from level 9 operators, which consist of $\mJ$ only.

\begin{table}[htbp]
  \centering
\begin{tabular}{|c|c|c|c|c|c|c|c|c|c|c|}
  \cline{1-2}\cline{4-5}\cline{7-8}\cline{10-11}
  level            & operator && level & operator     && level & operator            && level            & operator \\\cline{1-2}\cline{4-5}\cline{7-8}\cline{10-11}
  2                & $T$      &&       & $\mB$, $\mD$ &&       & $\mE$, $\mH$, $\mI$ && \multirow{2}*{8} & $TT\mA$  \\\cline{1-2}\cline{5-5}\cline{8-8}\cline{11-11}
  \multirow{2}*{4} & $\mA$    && 6     & $T\mA$       && 8     & $T\mB$, $T\mD$      &&                  & $TTTT$   \\\cline{2-2}\cline{5-5}\cline{8-8}\cline{10-11}
                   & $TT$     &&       & $TTT$        &&       & $\mA\mA$            && 9                & $\mJ$    \\\cline{1-2}\cline{4-5}\cline{7-8}\cline{10-11}
\end{tabular}
  \caption{The holomorphic nonidentity quasiprimary operators to be considered in this paper for $\CFT^n$ and up to level 9. We have omitted the replica indices and their constraints, which can be easily figured out and can also be found in \cite{Chen:2013dxa}.}\label{vacconfn}
\end{table}

Each of the CFT$^n$ quasiprimary operator $\Phi_K$ in (\ref{z15}) has the form
\be
\Phi_K^{j_1,j_2,\cdots,j_k} = \mX_1^{j_1}\mX_2^{j_2} \cdots \mX_k^{j_k},
\ee
with $\mX_1$, $\mX_2$, $\cdots$, $\mX_k$ being nonidentity quasiprimary operators in table~\ref{vacconf} and there are also some constraints for the $k$ replica indices $j_1,j_2,\cdots,j_k$. We have the one-point functions that are independent of the replica indices
\be
\lag \Phi_K^{j_1j_2\cdots j_k} \rag_\mR = \lag \mX_1 \rag_\mR \lag \mX_2 \rag_\mR \cdots \lag \mX_k \rag_\mR,
\ee
and so we can define $b_K$ from the OPE coefficient $d_K^{j_1j_2\cdots j_k}$ by summing over the replica indices \cite{Chen:2016lbu}
\be \label{defbK}
b_K = \sum_{j_1,j_2,\cdots,j_k} d_K^{j_1j_2\cdots j_k} \textrm{~with~some~constraints~for~}0 \leq j_1,j_2,\cdots,j_k \leq n-1.
\ee
To level 8 the form of $b_K$ can be found in \cite{Chen:2016lbu,He:2017vyf}, and from (\ref{z16}) we know
\be \label{z17}
b_\mJ = 0.
\ee
Then we write (\ref{z21}) explicitly as
\bea \label{haha1}
&& \tr_A \r_A^n = \f{c_n}{\ell^{2h_\s}}
\Big[ 1+b_T\lag T\rag_\mR \ell^2
+ \big( b_\mA\lag\mA\rag_\mR
       +b_{TT}\lag T\rag_\mR^2 \big) \ell^4
+ \big( b_\mB\lag\mB\rag_\mR
      + b_\mD\lag\mD\rag_\mR \nn\\
&& \phantom{\tr_A \r_A^n =}
      + b_{T\mA}\lag T\rag_\mR\lag\mA\rag_\mR
      + b_{TTT}\lag T\rag_\mR^3 \big) \ell^6
+ \big( b_\mE\lag\mE\rag_\mR
      + b_\mH\lag\mH\rag_\mR
      + b_\mI\lag\mI\rag_\mR \nn\\
&& \phantom{\tr_A \r_A^n =}
      + b_{T\mB}\lag T\rag_\mR\lag\mB\rag_\mR
      + b_{T\mD}\lag T\rag_\mR\lag\mD\rag_\mR
      + b_{\mA\mA}\lag\mA\rag_\mR^2
      + b_{TT\mA}\lag T\rag_\mR^2\lag\mA\rag_\mR \nn\\
&& \phantom{\tr_A \r_A^n =}
      + b_{TTTT}\lag T\rag_\mR^4 \big) \ell^8 + O(\ell^{10})
\Big].
\eea
Due to the absence of level 9 contribution, in the above the unknown terms start from
 $O(\ell^{10})$.

\begin{figure}[htbp]\centering
\subfigure[]{\includegraphics[height=5cm]{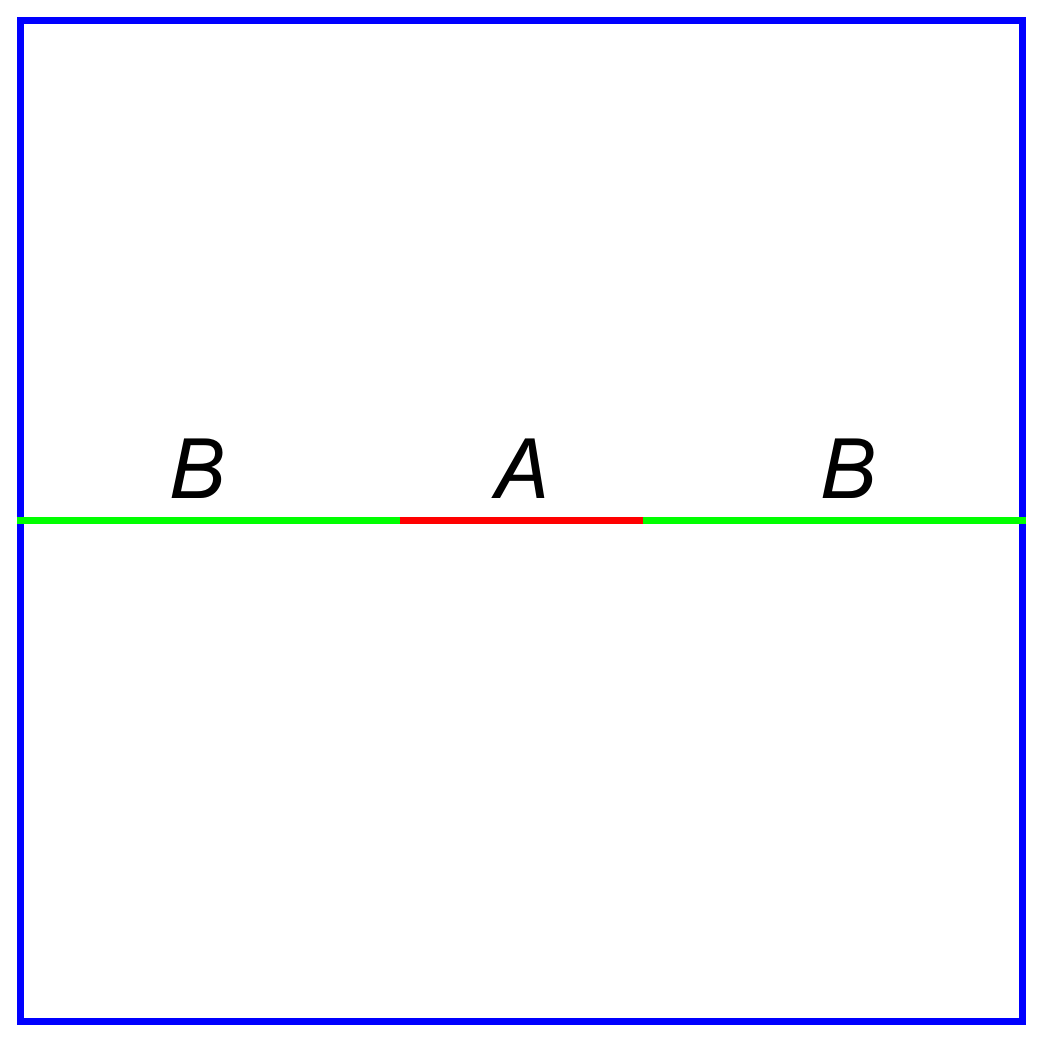}        \label{fig1a}} ~~
\subfigure[]{\includegraphics[height=5cm]{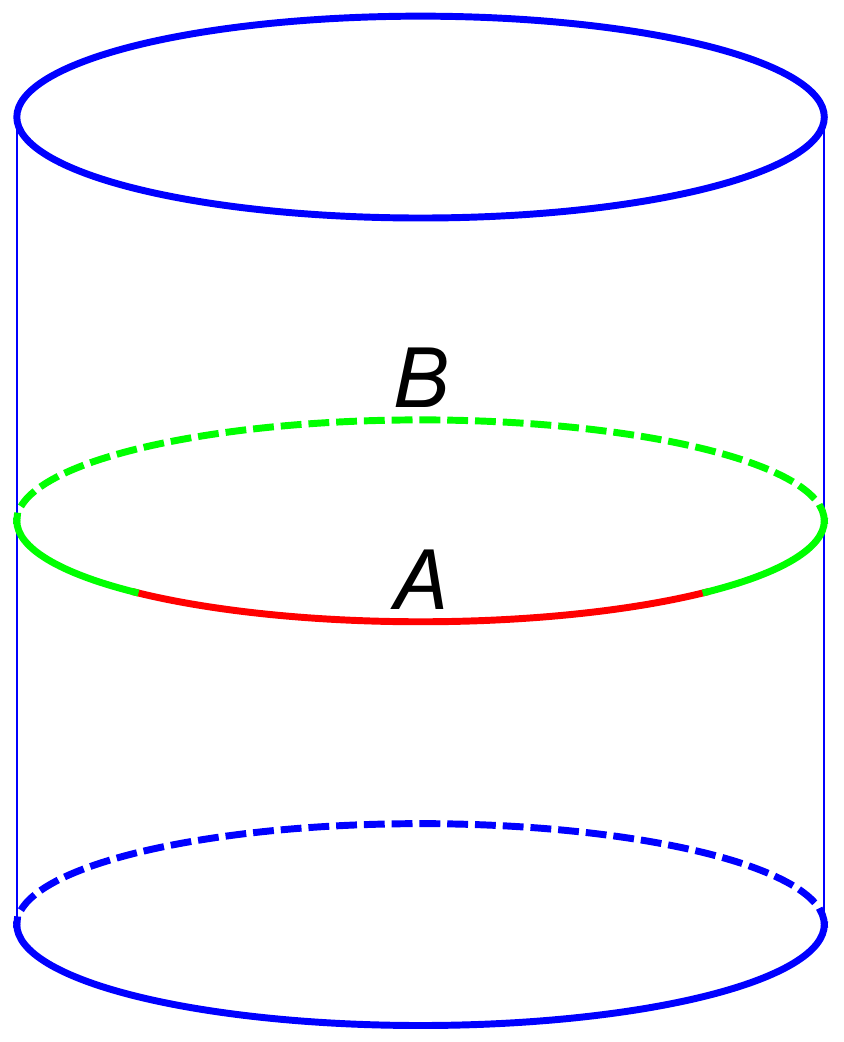}    \label{fig1b}} ~~
\subfigure[]{\includegraphics[height=5cm]{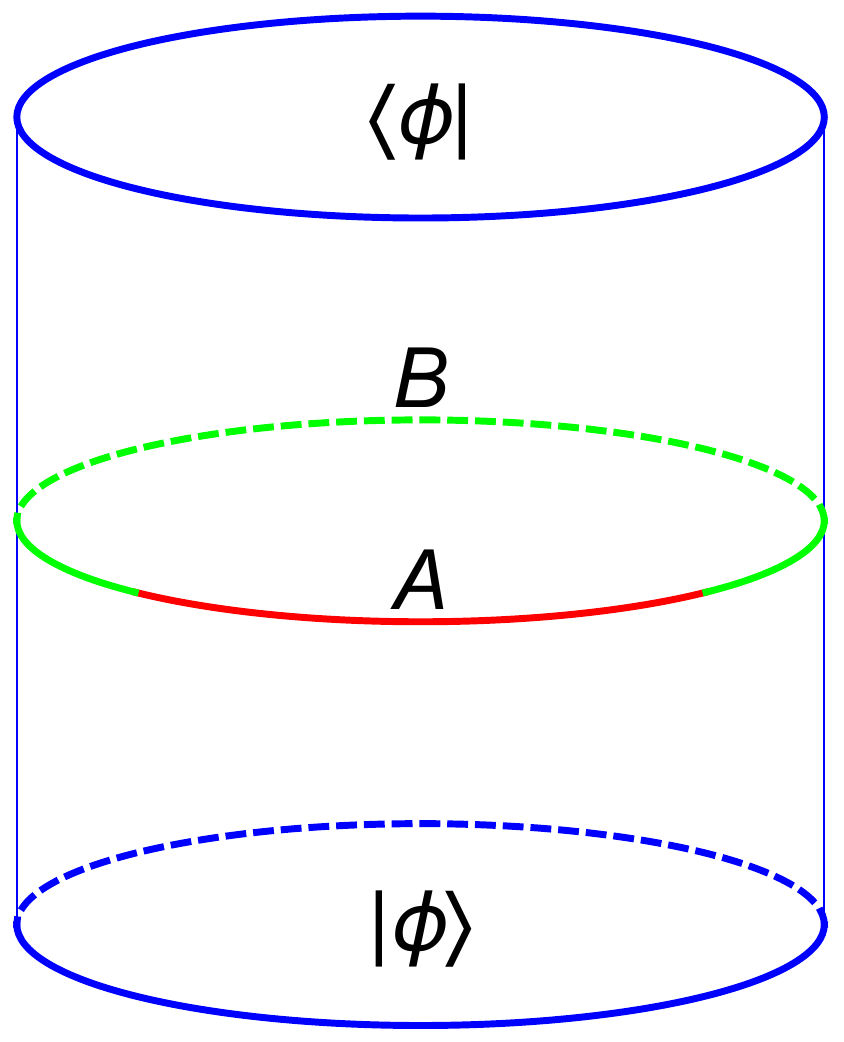} \label{fig1c}}\\
\subfigure[]{\includegraphics[width=5cm]{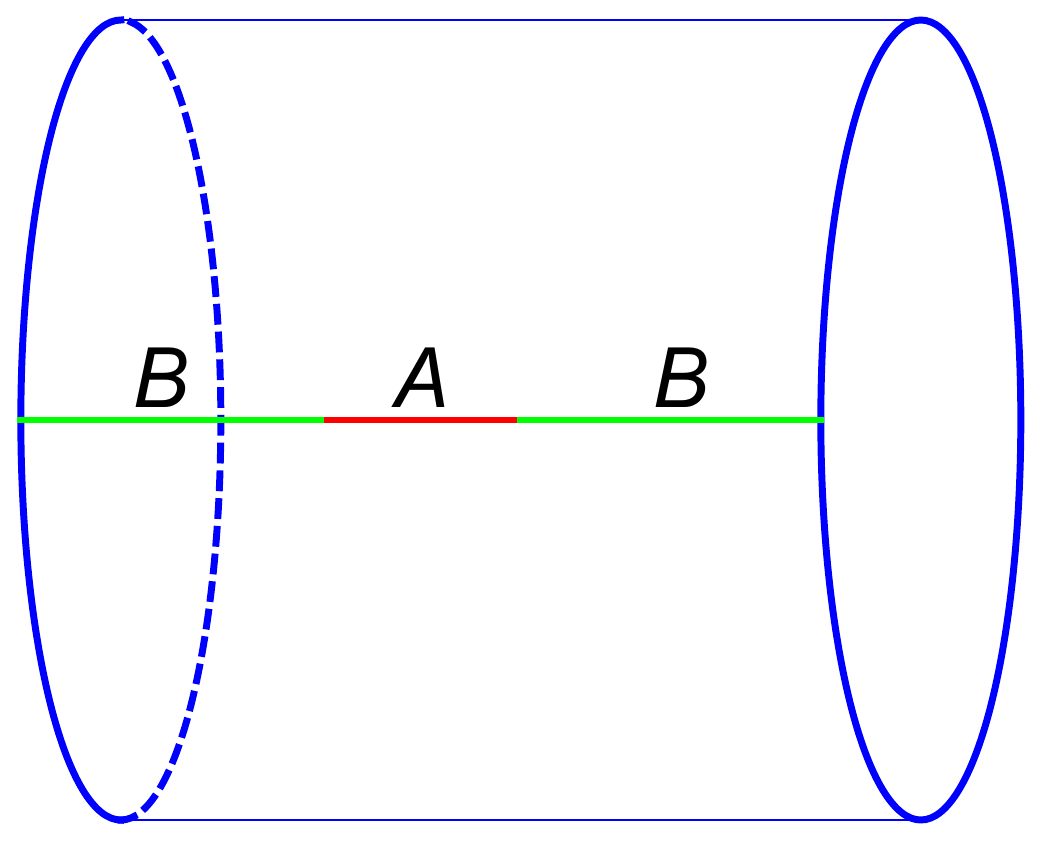}     \label{fig1d}} ~~
\subfigure[]{\includegraphics[width=5cm]{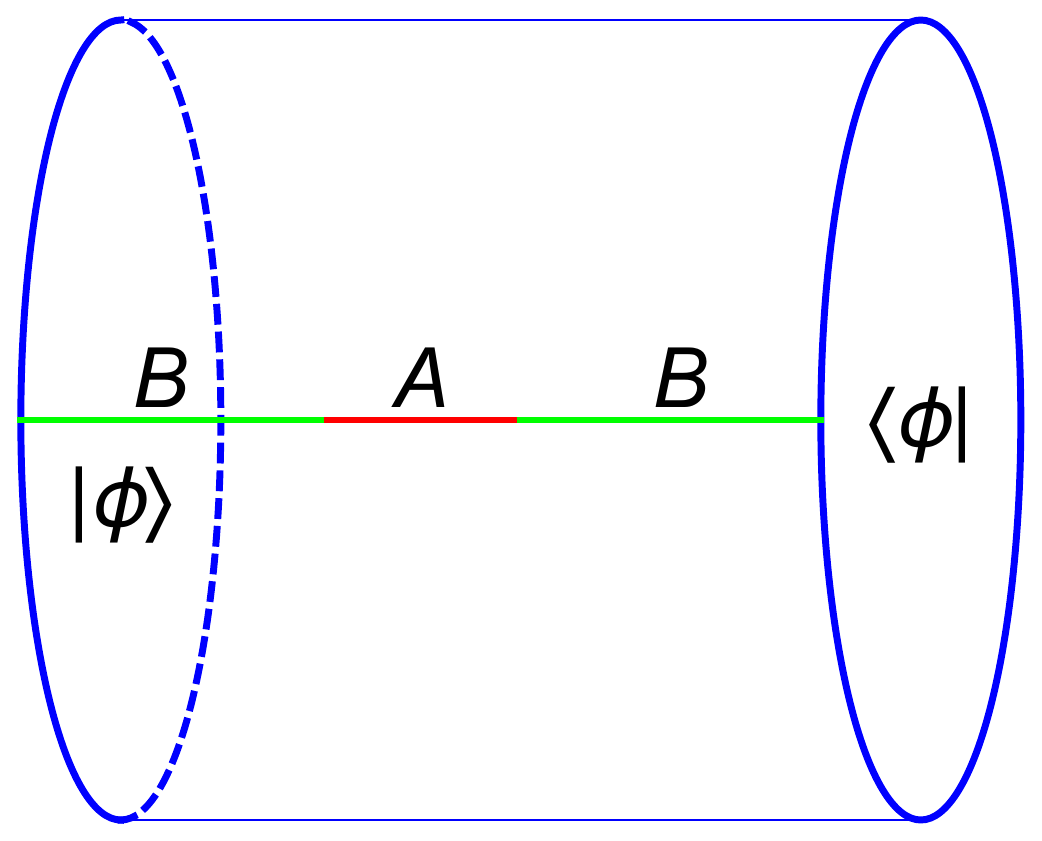}  \label{fig1e}}\\
\subfigure[]{\includegraphics[height=6cm]{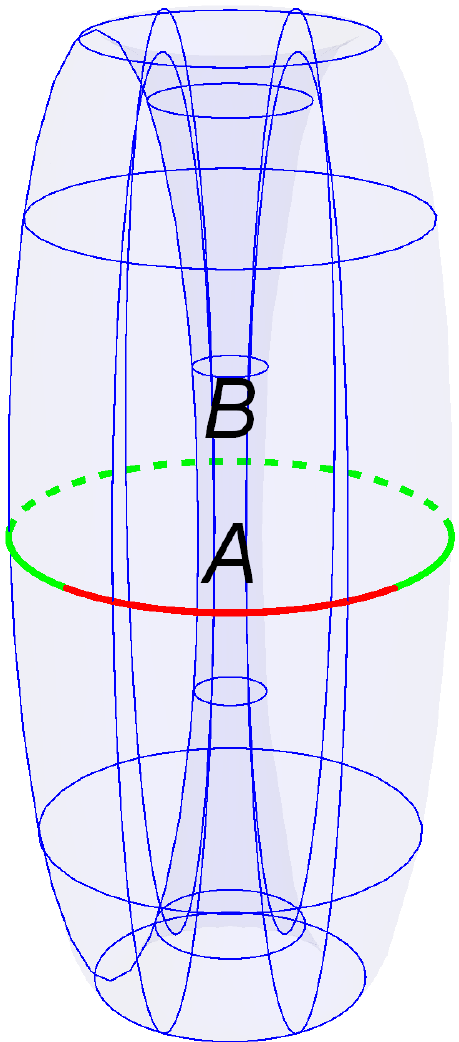}    \label{fig1f}} ~~
\subfigure[]{\includegraphics[height=6cm]{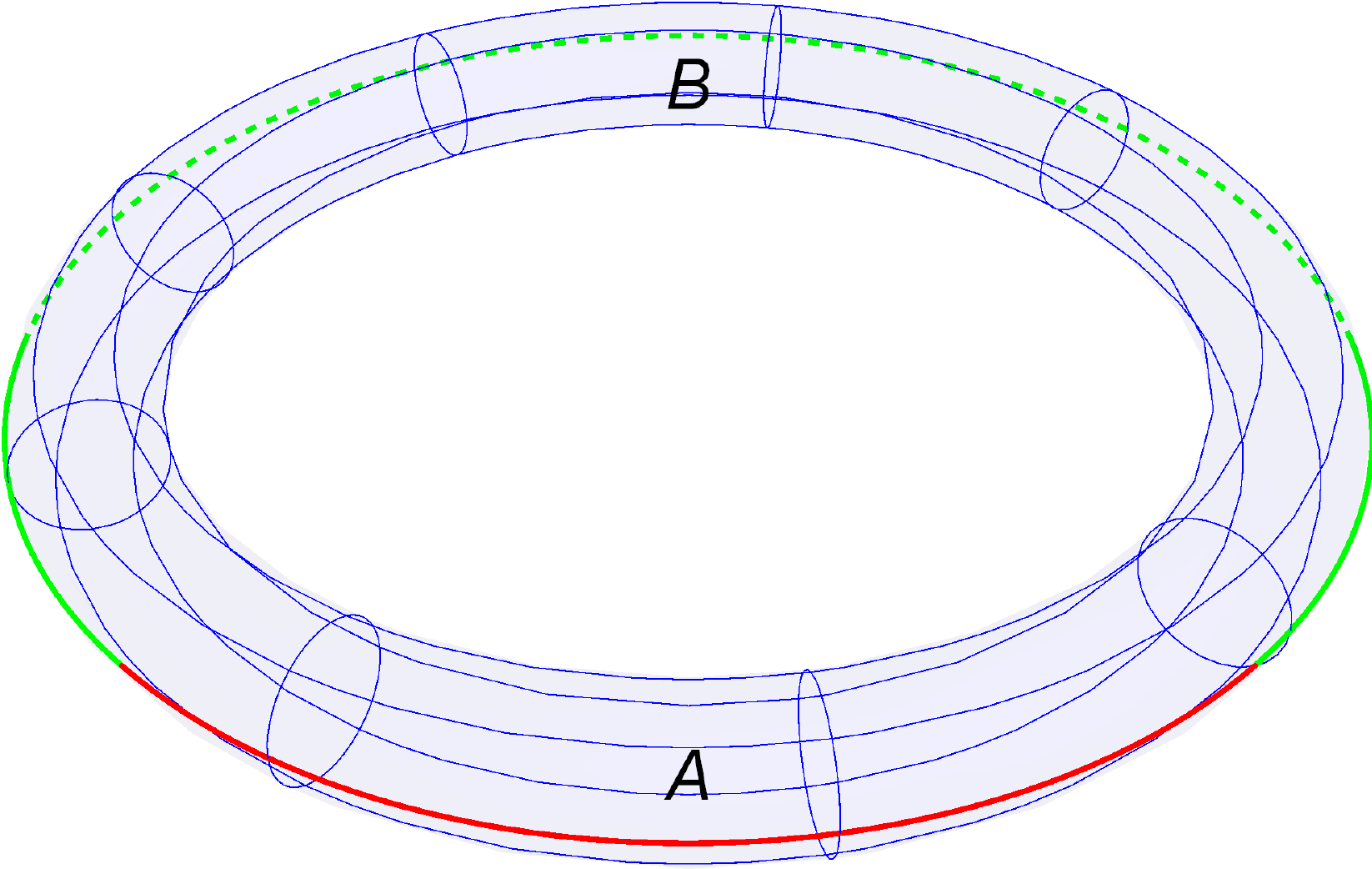}    \label{fig1g}}
\caption{The Riemann surfaces as environments for the interval $A=[0,\ell]$ we consider in this paper. (a) A complex plane $\mR(\emptyset)$. (b) A vertical cylinder $\mR(L)$. (c) A vertical cylinder capped with operators $\mR(L,\phi)$. (d) A horizontal cylinder $\mR(\b)$. (e) A horizontal cylinder capped with operators $\mR(\b,\phi)$. (f) A fat torus $\mR(L,q=\ep^{-2\pi\b/L})$. (g) A thin torus $\mR(\beta,p=\ep^{-2\pi L/\b})$.}\label{fig1}
\end{figure}

In this paper we consider several different Riemann surfaces that are environments of a short interval $A=[0,\ell]$, and they are shown in figure~\ref{fig1}. Note that the complex plane case figure~\ref{fig1a} can be got as limits of other six cases.
\begin{itemize}
  \item In figure~\ref{fig1a}, the interval is on an infinite straight line in ground state of the CFT. It is just a complex plane $\mR(\es)$, and we denote the total system density matrix as $\r(\es)$ and reduced density matrix as $\r_A(\es)$.
  \item In figure~\ref{fig1b}, the interval is on a length $L$ circle in ground state, and it is a vertical cylinder $\mR(L)$. We have the density matrix $\r(L)$ and reduced density matrix $\r_A(L)$.
  \item In figure~\ref{fig1c}, the interval is on a circle in excited state $|\phi\rag$ of a primary operator $\phi$ with conformal weight $h_\phi$ and normalization $\a_\phi=1$. The manifold is a vertical cylinder capped with an operator inserted at each of the two ends, and we denote it as $\mR(L,\phi)$. We have the density matrix $\r(L,\phi)$ and reduced density matrix $\r_A(L,\phi)$.
  \item In figure~\ref{fig1d}, the interval is on an infinite straight line in thermal state with inverse temperature $\b$. The manifold is a horizontal cylinder $\mR(\b)$, and it is the modular transformation of $\mR(L)$. We have the density matrix $\r(\b)$ and reduced density matrix $\r_A(\b)$.
  \item Figure~\ref{fig1e} is the modular transformation of figure~\ref{fig1c}. The interval is on an infinite straight line in thermal state with inverse temperature $\b$, and also there are boundary conditions imposed on both ends of the horizontal cylinder. Each boundary condition is effectively represented by insertion of a primary operator $\phi$. We have the Riemann surface $\mR(\b,\phi)$, the density matrix $\r(\b,\phi)$ and reduced density matrix $\r_A(\b,\phi)$.
  \item In figure~\ref{fig1f}, the interval is on a length $L$ circle in thermal state with inverse temperature $\b$. The temperature is low $\b \gg L$, and the manifold is a fat torus. In limit $\b/L \to \inf$, it becomes a vertical cylinder figure~\ref{fig1b}. We have the Riemann surface $\mR(L,q)$, the density matrix $\r(L,q)$ and reduced density matrix $\r_A(L,q)$, with definition $q=\ep^{-2\pi\b/L}$.
  \item In figure~\ref{fig1g}, the interval is on a length $L$ circle in thermal state with inverse temperature $\b$. The temperature is high $L \gg \b$, the manifold is a thin torus, and it is the modular transformation of the fat torus figure~\ref{fig1f}. In limit $L/\b \to \inf$, it becomes the horizontal cylinder figure~\ref{fig1d}. We have the Riemann surface $\mR(\b,p)$, the density matrix $\r(\b,p)$ and reduced density matrix $\r_A(\b,p)$, with $p=\ep^{-2\pi L/\b}$.
\end{itemize}

We need the one-point functions $\lag \mX \rag_\mR$ with $\mX=T,\mA,\mB,\mD,\mE,\mH,\mI$ for $\mR$ being each of these Riemann surfaces in figure~\ref{fig1}. In practice, we only need to consider the cases of $\mR(L,\phi)$ and $\mR(L,q)$, and the other cases can be got from them by some simple substitutes and/or limits. For the case $\mR(L,\phi)$ one can find the results in \cite{He:2017vyf}. For the case $\mR(L,q)$ one can find the results to level 6 in \cite{Chen:2016lbu}. Using the method in appendix B of \cite{Chen:2016lbu}, the conformal transformations of $\mE,\mH,\mI$ in \cite{He:2017vyf}, as well as the structure constants in \cite{Chen:2016lbu} and (\ref{z18}), we get the one-point functions
\bea
&& \lag \mE \rag_{\mR(L,q)} =  \frac{23452 \pi^8 c}{59535 L^8}
                              +\frac{36608 \pi^8 (1008 c-1) q^2}{3969 L^8}
                              +\frac{18304 \pi^8 (13608 c+335) q^3}{1323 L^8}   \nn\\
&& \phantom{\lag \mE \rag_{\mR(L,q)} =}
                              +\frac{36608 \pi^8 (54096 c+5795) q^4}{1323L^8}
                              +O(q^5),   \nn\\
&& \lag \mH \rag_{\mR(L,q)} = -\frac{13 \pi^8 c (5 c+22) (465 c-127)}{10125(105 c+11)L^8}
                              +\frac{1664 \pi^8 (5 c+22) (945 c^2+2184 c-10) q^2}{675 (105 c+11)L^8}   \nn\\
&& \phantom{\lag \mE \rag_{\mR(L,q)} =}
                              +\frac{416 \pi^8 (5 c+22) (11340 c^2+31323 c-1517) q^3}{225 (105 c+11)L^8}   \\
&& \phantom{\lag \mE \rag_{\mR(L,q)} =}
                              +\frac{1664 \pi^8 (64575 c^3+334935 c^2+226879 c+26048) q^4}{225 (105 c+11)L^8}
                              +O(q^5),   \nn\\
&& \lag \mI \rag_{\mR(L,q)} = \frac{\pi^8 c (2 c-1) (3 c+46) (5 c+3) (5 c+22) (7 c+68)}{1296 (1050 c^2+3305c-251) L^8}
             \Big( 1+\frac{3264 q^2}{c}+\frac{13536 q^3}{c}   \nn\\
&& \phantom{\lag \mE \rag_{\mR(L,q)} =}
             +\frac{576 (325 c+4814) q^4}{c (5c+22)}+O(q^5) \Big).   \nn
\eea

\subsection{Partition function from twist operators}

Gluing $n$ reduced density matrices $\r_{A,j}$ on $n$ different Riemann surface $\mR_j$ with $j=0,1,\cdots,n-1$, one gets a CFT on the Riemann surface $\mR^n=\mR_0\oplus\cdots\oplus\mR_{n-1}$. This suggests to assume that the partition function on $\mR^n$ can still be written as a two-point function of twist operators
\be \label{z19}
\tr_A(\r_{A,0}\cdots\r_{A,n-1}) = \lag \mT(\ell)\td\mT(0)\rag_{\mR_0\oplus\cdots\oplus\mR_{n-1}}.
\ee
Each replica of the CFT lives on one of the Riemann surfaces, and different replicas are connected only by twist operators. For the $n=2$ and $n=3$ cases one can see, for examples, \cite{Gaberdiel:2010jf,Cardy:2014rqa,Mandal:2015jla,Headrick:2015gba,Cardy:2015xaa,Belin:2017nze,Basu:2017kzo,Cardy:2017qhl,Keller:2017iql,Cho:2017fzo}, but we are not sure if it is applicable for general $n$ when $\Z_n$ replica symmetry is lost. Actually, in this paper we only use a relaxed relation
\be \label{z20}
\f1{n!} [ \tr_A(\r_{A,0}\cdots\r_{A,n-1}) + {\textrm{permutations}} ]
= \f1{n!} [ \lag \mT(\ell)\td\mT(0)\rag_{\mR_0\oplus\cdots\oplus\mR_{n-1}} + {\textrm{permutations}} ],
\ee
and $\Z_n$ replica symmetry is recovered after permutations. Thus when we write (\ref{z19}), we actually mean (\ref{z20}), and there is caveat that (\ref{z20}) basically is an assumption that we have no concrete proof.

For two different Riemann surfaces $\mR$ and $\mS$, we may define respectively two reduced density matrices $\r_A$ and $\s_A$. In this paper, we need to calculate the partition function
\be
\tr_A (\r_A^m\s_A^{n-m} ),
\ee
with $n$ being an integer and $m=0,1,\cdots,n$.
Using (\ref{z20}), we see that it is just the right-hand side of (\ref{haha1}) with the substitutes of the forms
\bea
&& b_\mX \lag \mX \rag_\mR \to \f{b_\mX}{n} \big[ m \lag \mX \rag_\mR + (n-m)\lag \mX \rag_\mS \big], \nn\\
&& b_{\mX\mX} \lag \mX \rag_\mR^2 \to \f{b_{\mX\mX}}{n(n-1)}
                                      \big[ m(m-1) \lag \mX \rag_\mR^2
                                      + 2m(n-m)\lag \mX \rag_\mR\lag \mX \rag_\mS
                                      + (n-m)(n-m-1) \lag \mX \rag_\mS^2 \big], \nn\\
&& b_{\mX\mY} \lag \mX \rag_\mR \lag \mY \rag_\mR \to \f{b_{\mX\mY}}{n(n-1)}
                                                     \big[ m(m-1) \lag \mX \rag_\mR\lag \mY \rag_\mR
                                                     + m(n-m)(\lag\mX \rag_\mR\lag \mY \rag_\mS + \lag\mX\rag_\mS\lag \mY \rag_\mR) \nn\\
&& \phantom{b_{\mX\mY} \lag \mX \rag_\mR \lag \mY \rag_\mR \to}
                                                     + (n-m)(n-m-1) \lag \mX \rag_\mS \lag \mY \rag_\mS \big], \nn\\
&& b_{\mX\mX\mX} \lag \mX \rag_\mR^3 \to \f{b_{\mX\mX\mX}}{n(n-1)(n-2)}
                                         \big[ m(m-1)(m-2) \lag \mX \rag_\mR^3
                                         + 3m(m-1)(n-m)\lag \mX \rag_\mR^2\lag \mX \rag_\mS \nn\\
&& \phantom{b_{\mX\mX\mX} \lag \mX \rag_\mR^3 \to}
                                         + 3m(n-m)(n-m-1)\lag \mX \rag_\mR \lag \mX \rag_\mS^2
                                         + (n-m)(n-m-1)(n-m-2) \lag \mX \rag_\mS^3 \big], \nn\\
&& b_{\mX\mX\mY} \lag \mX \rag_\mR^2\lag \mY \rag_\mR \to \f{b_{\mX\mX\mY}}{n(n-1)(n-2)}
                                         \big[ m(m-1)(m-2) \lag \mX \rag_\mR^2\lag \mY \rag_\mR \nn\\
&& \phantom{b_{\mX\mX\mY} \lag \mX \rag_\mR^2\lag \mY \rag_\mR \to}
                                         + m(m-1)(n-m)(\lag \mX \rag_\mR^2\lag \mY \rag_\mS
                                                       + 2 \lag \mX \rag_\mR \lag \mX \rag_\mS \lag \mY \rag_\mR)\\
&& \phantom{b_{\mX\mX\mY} \lag \mX \rag_\mR^2\lag \mY \rag_\mR \to}
                                         + m(n-m)(n-m-1)(2 \lag \mX \rag_\mR \lag\mX \rag_\mS \lag \mY \rag_\mS
                                                         + \lag \mX \rag_\mS^2 \lag \mY \rag_\mR) \nn\\
&& \phantom{b_{\mX\mX\mY} \lag \mX \rag_\mR^2\lag \mY \rag_\mR \to}
                                         + (n-m)(n-m-1)(n-m-2) \lag \mX \rag_\mS^2\lag \mY \rag_\mS \big], \nn\\
&& b_{\mX\mX\mX\mX} \lag \mX \rag_\mR^4 \to \f{b_{\mX\mX\mX\mX}}{n(n-1)(n-2)(n-3)}
                                         \big[ m(m-1)(m-2)(m-3) \lag \mX \rag_\mR^4 \nn\\
&& \phantom{b_{\mX\mX\mX\mX} \lag \mX \rag_\mR^4 \to}
                                         + 4m(m-1)(m-2)(n-m)\lag \mX \rag_\mR^3\lag \mX \rag_\mS \nn\\
&& \phantom{b_{\mX\mX\mX\mX} \lag \mX \rag_\mR^4 \to}
                                         + 6m(m-1)(n-m)(n-m-1)\lag \mX \rag_\mR^2 \lag \mX \rag_\mS^2\nn\\
&& \phantom{b_{\mX\mX\mX\mX} \lag \mX \rag_\mR^4 \to}
                                         + 4m(n-m)(n-m-1)(n-m-2) \lag \mX \rag_\mR\lag \mX \rag_\mS^3\nn\\
&& \phantom{b_{\mX\mX\mX\mX} \lag \mX \rag_\mR^4 \to}
                                         + (n-m)(n-m-1)(n-m-2)(n-m-3) \lag \mX \rag_\mS^4 \big], \nn
\eea
with $\mX$, $\mY$ denoting general quasiprimary operators. A general substitute takes the form
\be
b_{\mX_1\mX_2\cdots\mX_k} \lag \mX_1 \rag_\mR \lag \mX_2 \rag_\mR \cdots \lag \mX_k \rag_\mR
\to
\f{b_{\mX_1\mX_2\cdots\mX_k}}{C_n^k} \big( C_m^k \lag\mX_1 \rag_\mR \lag \mX_2 \rag_\mR \cdots \lag \mX_k \rag_\mR + \cdots  \big),
\ee
with $C_n^k$ and $C_m^k$ being the binomial coefficients, and in the right hand side we have omitted various terms with some $\mR$'s being replaced by $\mS$'s.

In section~\ref{secjsd}, we need to calculate the partition function
\be \label{z28}
\tr_A\Big(\f{\r_A+\s_A}{2}\Big)^n = \f1{2^n} \sum_{m=0}^n C_n^m \tr_A (\r_A^m\s_A^{n-m} ),
\ee
with $\tr_A (\r_A^m\s_A^{n-m} )$ being understood as the left-hand side of (\ref{z20}).
Using the summation formulas
\bea \label{z50}
&& \sum_{m=0}^n C_n^m m =2^{n-1} n, ~~
   \sum_{m=0}^n C_n^m m(m-1)=\sum_{m=0}^n C_n^m m(n-m)=2^{n-2} n(n-1), \nn\\
&& \sum_{m=0}^n C_n^m m(m-1)(m-2)=\sum_{m=0}^n C_n^m m(m-1)(n-m)=2^{n-3} n(n-1)(n-2), \nn\\
&& \sum_{m=0}^n C_n^m m(m-1)(m-2)(m-3)
  =\sum_{m=0}^n C_n^m m(m-1)(m-2)(n-m)\\
&&=\sum_{m=0}^n C_n^m m(m-1)(n-m)(n-m-1)
  =2^{n-4} n(n-1)(n-2)(n-3),\nn
\eea
we get that (\ref{z28}) is just the right-hand side of (\ref{haha1}) with the substitutes
\bea \label{z43}
&& b_\mX \lag \mX \rag_\mR \to \f{b_\mX}{2} \big( \lag \mX \rag_\mR + \lag \mX \rag_\mS \big), \nn\\
&& b_{\mX\mX} \lag \mX \rag_\mR^2 \to \f{b_{\mX\mX}}{4}
                                      \big( \lag \mX \rag_\mR + \lag \mX \rag_\mS \big)^2, \nn\\
&& b_{\mX\mY} \lag \mX \rag_\mR \lag \mY \rag_\mR \to \f{b_{\mX\mY}}{4}
                                                     \big( \lag \mX \rag_\mR + \lag \mX \rag_\mS \big)
                                                     \big( \lag \mY \rag_\mR + \lag \mY \rag_\mS \big), \nn\\
&& b_{\mX\mX\mX} \lag \mX \rag_\mR^3 \to \f{b_{\mX\mX\mX}}{8}
                                         \big( \lag \mX \rag_\mR + \lag \mX \rag_\mS \big)^3, \\
&& b_{\mX\mX\mY} \lag \mX \rag_\mR^2\lag \mY \rag_\mR \to \f{b_{\mX\mX\mY}}{8}
                                         \big( \lag \mX \rag_\mR + \lag \mX \rag_\mS \big)^2
                                         \big( \lag \mY \rag_\mR + \lag \mY \rag_\mS \big), \nn\\
&& b_{\mX\mX\mX\mX} \lag \mX \rag_\mR^4 \to \f{b_{\mX\mX\mX\mX}}{16}
                                         \big( \lag \mX \rag_\mR + \lag \mX \rag_\mS \big)^4. \nn
\eea

In section~\ref{secs24}, we need to calculate
\be
\tr_A(\r_A-\s_A)^n = \sum_{m=0}^n C_n^m (-)^{n-m} \tr_A (\r_A^m\s_A^{n-m}).
\ee
Using the fact that
\bea
&& \sum_{m=0}^n C_n^m (-)^{n-m} m^k =0 {\rm{~for~}} k=0,1,\cdots,n-1, \nn\\
&& \sum_{m=0}^n C_n^m (-)^{n-m} m^n = n!,
\eea
we get
\be \label{z32}
\tr_A(\r_A-\s_A)^n = \sum_{\{\mX_1,\mX_2,\cdots,\mX_n\}} b_{\mX_1\mX_2\cdots\mX_n}
\big(\lag \mX_1 \rag_\mR - \lag \mX_1 \rag_\mS\big)
\big(\lag \mX_2 \rag_\mR - \lag \mX_2 \rag_\mS\big)
\cdots
\big(\lag \mX_n \rag_\mR - \lag \mX_n \rag_\mS\big).
\ee
Note that the summation of $\{\mX_1,\mX_2,\cdots,\mX_n\}$ is over different sets of nonidentity quasiprimary operators and the order of the operators in each set does not matter.
For $n=2$ it is just the result in \cite{Basu:2017kzo}. Note that for general $n$, $b_{\mX_1\mX_2\cdots\mX_n}$ is complex and has no universal form, and it is related to the $n$-point correlation function on complex plane
$\lag \mX_1(z_1)\mX_2(z_2) \cdots \mX_n(z_n) \rag_\mC$.

\subsection{The $n\to1$ limit}\label{sec2.4}

If we are only interested in the $n\to1$ limit
\be
-\f{\log\tr_A \r_A^n }{n-1}\Big|_{n\to1},
\ee
instead of the general $n$ result, there can be a simpler calculation \cite{Beccaria:2014lqa,Li:2016pwu}. For each CFT$^n$ operator $\Phi_K$, we may define
\be
a_K = - \lim_{n \to 1} \f{b_K}{n-1},
\ee
with $b_K$ being defined in (\ref{defbK}). Using the results of $b_K$ in \cite{Chen:2016lbu,He:2017vyf}, we get the relevant results of $a_K$
\bea
&& a_T = -\frac{1}{6}, ~~
   a_{TT} = -\frac{1}{30 c}, ~~
   a_{TTT} = -\frac{4}{315 c^2}, \\
&& a_{\mA\mA} = -\frac{1}{126 c (5 c+22)}, ~~
   a_{TT\mA} = \frac{1}{315 c^2}, ~~
   a_{TTTT} = -\frac{c+8}{630 c^3}. \nn
\eea
For the reduced density matrix $\r_A$ on Riemann surface $\mR$, we get
\bea \label{z42}
&& -\f{\log\tr_A \r_A^n }{n-1}\Big|_{n\to1} =
\f{c}{6}\log\f{\ell}{\e} + a_T \lag T \rag_\mR \ell^2 + a_{TT} \lag T \rag_\mR^2 \ell^4
+ a_{TTT} \lag T \rag_\mR^3 \ell^6
+ \big( a_{\mA\mA} \lag \mA \rag_\mR^2 \nn\\
&& \phantom{-\f{\log\tr_A \r_A^n }{n-1}\Big|_{n\to1} =}
      + a_{TT\mA} \lag T \rag_\mR^2\lag \mA \rag_\mR
      + a_{TTTT} \lag T \rag_\mR^4 \big) \ell^8 + O(\ell^{10}).
\eea
For the reduced density matrix $\r_A$, $\s_A$, defined respectively on Riemann surface $\mR$, $\mS$, we get that
\be
-\f{\log\tr_A (\r_A \s_A^{n-1}) }{n-1}\Big|_{n\to1}
\ee
equals right-hand side of (\ref{z42}) with the substitutes
\bea
&& a_{\mX\mX} \lag \mX \rag_\mR^2 \to a_{\mX\mX}\big( 2\lag \mX \rag_\mR-\lag \mX \rag_\mS\big)\lag \mX \rag_\mS, \nn\\
&& a_{\mX\mX\mX} \lag \mX \rag_\mR^3 \to a_{\mX\mX\mX}\big(3\lag \mX \rag_\mR-2\lag \mX \rag_\mS\big)\lag \mX \rag_\mS^2, \\
&& a_{\mX\mX\mY} \lag \mX \rag_\mR^2\lag \mY \rag_\mR \to a_{\mX\mX\mY}
        \big( 2 \lag \mX \rag_\mR \lag \mX \rag_\mS \lag \mY \rag_\mS
          + \lag \mX \rag_\mS^2 \lag \mY \rag_\mR
          -2 \lag \mX \rag_\mS^2\lag \mY \rag_\mS \big), \nn\\
&& a_{\mX\mX\mX\mX} \lag \mX \rag_\mR^4 \to a_{\mX\mX\mX\mX}\big(4\lag \mX \rag_\mR-3\lag \mX \rag_\mS\big)\lag \mX \rag_\mS^3. \nn
\eea
Similarly, we get that
\be
-\f{\log\tr_A (\f{\r_A+\s_A}{2})^n }{n-1}\Big|_{n\to1}
\ee
equals right-hand side of (\ref{z42}) with the substitutes (\ref{z43}).

\subsection{R\'enyi and entanglement entropies on various Riemann surfaces}\label{secren}
Using the above prescriptions, we can evaluate the entanglement and R\'enyi entropies on various Riemann surfaces, some of which {have} been obtained before. The results will then serve in the next section for calculating the dissimilarity measures between reduced density matrices.

For a reduced density matrix $\r_A$, the R\'enyi entropy is defined as
\be
S_n = -\f{1}{n-1} \log \tr_A \r_A^n,
\ee
and taking the $n\to1$ limit one can get the entanglement entropy
\be\label{SEE}
S = - \tr_A ( \r_A \log \r_A ).
\ee
The R\'enyi entropy can be calculated from (\ref{haha1}), and the entanglement entropy can be calculated from the $n\to1$ limit of the R\'enyi entropy or directly from (\ref{z42}).

\begin{figure}[htbp]
  \centering
  \includegraphics[width=0.4\textwidth]{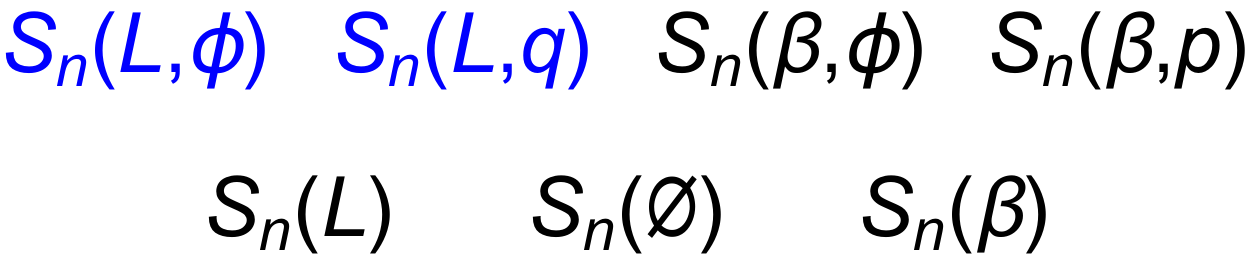}\\
  \caption{The seven R\'enyi entropies we can calculate using OPE of the twist operators. In practice, we only need to calculate $S_n(L,\phi)$ and $S_n(L,\phi)$, as marked in blue, and the other cases can be obtained easily from them.}\label{ren}
\end{figure}

We calculate the R\'enyi entropies and entanglement entropies for the seven Riemann surfaces in figure~\ref{fig1}. The seven R\'enyi entropies are shown in figure~\ref{ren}. In practice we only need to calculate $S_n(L,\phi)$ and $S_n(L,q)$, as marked in blue in the figure, and the other cases can be obtained easily from them. Note that most of the results in this section are not new, and just serves as a check of the OPE coefficients and the one-point functions.

R\'enyi entropy $S_n(L,\phi)$ and entanglement entropy $S(L,\phi)$ have been calculated in \cite{He:2017vyf}\footnote{One can also follow different approach \cite{He:2014mwa,Guo:2015uwa,Chen:2015usa} to obtain R\'enyi entropy $S_n(L,\phi)$ with finite size system in 2D rational CFTs.}, and we will not repeat the results here. Since now at level 9 we have (\ref{z17}), the unknown terms $O(\ell^9)$ in results of \cite{He:2017vyf} are actually of order $O(\ell^{10})$. For the reduced density matrix $\r_A(\b,\phi)$, we have the R\'enyi entropy and entanglement entropy
\be
S_n(\b,\phi) = S_n(L,\phi)|_{L \to \ii \b}, ~~ S(\b,\phi) = S(L,\phi)|_{L \to \ii \b}.
\ee

For $\r_A(L,q)$, the R\'enyi entropy and entanglement entropy have been calculated using OPE of the twist operators to order $\ell^7$ in \cite{Chen:2016lbu}, and here we calculate the results to order $\ell^9$. In large $c$ limit we write the R\'enyi entropy as the leading part, the next-to-leading part, the next-to-next-to-leading part, and etc,
\be
S_n(L,q) = S_n^\cL(L,q) + S_n^\NL(L,q) + S_n^\NNL(L,q) + \cdots,
\ee
and to order $\ell^9$ only the first three parts are non-vanishing. Explicitly, we have the leading part
\bea
&& S_n^\cL(L,q) = \frac{c (n+1)}{12 n} \log\f{\ell}{\e}
-\frac{\pi^2 c (n+1)}{72 n} \f{\ell^2}{L^2}
+ \Big( -\frac{\pi^4 c (n+1) }{2160 n}
        -\frac{\pi^4 c (n-1) (n+1)^2 q^2}{18 n^3} \nn\\
&& ~~~
        -\frac{2 \pi^4c (n-1) (n+1)^2q^3}{9 n^3}
        -\frac{11 \pi^4 c (n-1) (n+1)^2 q^4}{18 n^3} +O(q^5) \Big) \f{\ell^4}{L^4}\nn\\
&& ~~~
+ \Big( -\frac{\pi^6 c (n+1) }{34020 n}
        +\frac{\pi^6 c (n-1) (n+1)^2 q^2}{27 n^3}
        +\frac{10 \pi^6 c (n-1) (n+1)^2 q^3}{27 n^3}\nn\\
&& ~~~
        +\frac{\pi^6 c (n+1) (48 n^4-49 n^2+1) q^4}{27 n^5} + O(q^5) \Big) \f{\ell^6}{L^6}
+ \Big( -\frac{\pi^8 c (n+1) }{453600  n}\nn\\
&& ~~~
        -\frac{\pi^8 c (n-1) (n+1)^2 q^2}{90  n^3}
        -\frac{4 \pi^8 c (n-1) (n+1)^2 q^3}{15 n^3}\\
&& ~~~
        -\frac{\pi^8 c (n-1) (n+1)^2 (3727 n^4-62 n^2-11) q^4}{1620 n^7}
        +O(q^5) \Big) \f{\ell^8}{L^8}
+O(\ell^{10}) \nn
\eea
the next-to-leading part
\bea
&& \hspace{-8mm}
   S_n^\NL(L,q)= \Big( \frac{2\pi^2 (n+1) q^2}{3 n}
                      +\frac{\pi^2(n+1) q^3}{n}
                      +\frac{2\pi^2 (n+1) q^4}{n}+O(q^5) \Big) \f{\ell^2}{L^2} \nn\\
&& \hspace{-8mm}  ~~
+\Big( -\frac{\pi^4(n+1) (9 n^2-11) q^2}{45 n^3}
       -\frac{\pi^4(n+1) (41 n^2-44) q^3}{45 n^3}\nn\\
&& \hspace{-8mm}  ~~
       -\frac{\pi^4(n+1) (49 n^2-51) q^4}{15 n^3}
       +O(q^5) \Big)  \f{\ell^4}{L^4}
+\Big( \frac{2\pi^6 (n+1) (17 n^4-46 n^2+31) q^2}{945 n^5}\nn\\
&& \hspace{-8mm}  ~~
      +\frac{\pi^6 (n+1) (492n^4-1013 n^2+527) q^3}{945 n^5}
      +\frac{2\pi^6 (n+1) (1654 n^4-2903 n^2+1255) q^4}{945n^5}\nn\\
&& \hspace{-8mm}  ~~
      +O(q^5) \Big) \f{\ell^6}{L^6}
+\Big( -\frac{\pi^8 (n+1) (62 n^6-278 n^4+415 n^2-205) q^2}{14175 n^7}\\
&& \hspace{-8mm}  ~~
       -\frac{\pi^8 (n+1) (866 n^6-2694 n^4+2850 n^2-1025) q^3}{4725 n^7}\nn\\
&& \hspace{-8mm}  ~~
       -\frac{ \pi^8 (n+1)(66439 n^6-163681 n^4+139223 n^2-36257) q^4}{28350 n^7}
       +O(q^5) \Big) \f{\ell^8}{L^8}
       +O(\ell^{10}),\nn
\eea
and the next-to-next-to-leading part
\bea
&& S_n^\NNL(L,q) = \Big( -\frac{4 \pi^4(n+1) (n^2+11) q^4}{45 n^3 c}+O(q^5) \Big) \f{\ell^4}{L^4}\nn\\
&& ~~~
+\Big( \frac{4\pi^6 (n+1)( 26 n^4+271 n^2-345) q^4}{945 n^5 c}+O(q^5)\Big) \f{\ell^6}{L^6}\\
&& ~~~
+\Big(-\frac{8\pi^8 (n+1)(116 n^6+1141 n^4-3017 n^2+3398) q^4}{14175 n^7 c}+O(q^5) \Big) \f{\ell^8}{L^8}
   +O(\ell^{10}).\nn
\eea
The leading and next-to-leading parts match the results in \cite{Cardy:2014jwa,Chen:2014unl,Chen:2015uia}, which are calculated in another method. The $\ell^8$ order of the next-to-next-to-leading part is a new result.
Taking $n\to1$ limit we get the entanglement entropy
\bea
&& S(L,q) =
   \frac{c}{6} \log \frac{\ell }{\epsilon}
  +\Big( -\frac{c \pi^2}{36}+\frac{4 \pi^2 q^2}{3}+{2 \pi^2 q^3}+{4 \pi^2 q^4}+O(q^5) \Big)\f{\ell^2}{L^2} \nn\\
&& ~~~
  +\Big( -\frac{c \pi^4}{1080}+\frac{4 \pi^4 q^2}{45}+\frac{2 \pi^4 q^3}{15}+\frac{4 (c-8)\pi^4 q^4}{15 c}+O(q^5) \Big)\f{\ell^4}{L^4} \nn\\
&& ~~~
  +\Big(-\frac{c \pi^6}{17010}+\frac{8 \pi^6 q^2}{945}+\frac{4 \pi^6 q^3}{315}+\frac{8 (c-16) \pi^6 q^4}{315 c}+O(q^5) \Big)\f{\ell^6}{L^6}\\
&& ~~~
+\Big(-\frac{c \pi^8}{226800}
      +\frac{4 \pi^8q^2}{4725}
      +\frac{2 \pi^8 q^3}{1575}
      -\frac{4\pi^8(159 c+728) q^4}{1575 c}
      +O(q^5) \Big) \f{\ell^8}{L^8}
   +O(\ell^{10}).\nn
\eea

The R\'enyi entropy and entanglement entropy for $\r_A(\b,p)$ are just the modular transformation of those for $\r_A(L,q)$, i.e.,
\be
S_n(\b,p) = S_n(L,q)|_{L\to\ii\b,q\to p}, ~~
S(\b,p) = S(L,q)|_{L\to\ii\b,q\to p}.
\ee
Without considering the subtlety of boundary conditions at the entangling surface \cite{Ohmori:2014eia,Cardy:2016fqc}, the R\'enyi entropy and entanglement entropy for $\r_A(\es)$, $\r_A(L)$ and $\r_A(\b)$ are of universal forms and depend only on the central charge \cite{Calabrese:2004eu}
\bea
&& S_n(\es) = \f{c(n+1)}{12n} \log \f{\ell}{\e}, ~~
   S(\es) = \f{c}{6} \log \f{\ell}{\e}, \nn\\
&& S_n(L) = \f{c(n+1)}{12n} \log \Big( \f{L}{\pi\e}\sin\f{\pi\ell}{L} \Big), ~~
   S(L) = \f{c}{6} \log \Big( \f{L}{\pi\e}\sin\f{\pi\ell}{L} \Big), \nn\\
&& S_n(\b) = \f{c(n+1)}{12n} \log \Big( \f{\b}{\pi\e}\sinh\f{\pi\ell}{\b} \Big), ~~
   S(\b) = \f{c}{6} \log \Big( \f{\b}{\pi\e}\sinh\f{\pi\ell}{\b} \Big).
\eea
To order $\ell^9$ the above results can be obtained easily as the limits and/or substitutes of  $S_n(L,q)$, $S(L,q)$.

\section{Dissimilarities of reduced density matrices}\label{secmeasure}

In this section we evaluate various dissimilarity measures between reduced density matrices, which include relative entropy, Jensen-Shannon divergence, Schatten 2-norm and 4-norm. Some lengthy and not so enlightening results are collected in appendix~\ref{appcl} and the attached Mathematica notebook in arXiv.

\subsection{Relative entropy}\label{secrel}

The relative entropy is also called Kullback-Leibler divergence. For two reduced density matrices $\r_A$ and $\s_A$, the relative entropy is defined as
\be\label{REE}
S(\r_A\|\s_A) = \tr_A (\r_A\log\r_A) - \tr_A (\r_A\log\s_A).
\ee
To calculate the relative entropy, one may first calculate the $n$-th relative entropy
\be
S_n(\r_A\|\s_A) = \f1{n-1} [ \log \tr_A\r_A^n - \log \tr_A(\r_A\s_A^{n-1}) ],
\ee
and then takes the $n\to 1$ limit. The relative entropy is not symmetric for its two arguments, and one may define the symmetrized relative entropy
\be
S(\r_A,\s_A) = S(\r_A\|\s_A) + S(\s_A\|\r_A).
\ee
To calculate the symmetrized relative entropy, one can first calculate the $n$-th symmetrized relative entropy
\be
S_n(\r_A,\s_A) = S_n(\r_A\|\s_A) + S_n(\s_A\|\r_A),
\ee
and then takes the $n\to 1$ limit. It turns out that
\bea
&& S(\r_A \| \s_A) = -a_{TT}\big( \lag T \rag_\mR - \lag T \rag_\mS \big)^2 \ell^4
                     -a_{TTT}\big( \lag T \rag_\mR - \lag T \rag_\mS \big)^2 \big( \lag T \rag_\mR + 2 \lag T \rag_\mS \big) \ell^6 \nn\\
&& ~~~
  -\big[
        a_{\mA\mA}\big( \lag \mA \rag_\mR - \lag \mA \rag_\mS \big)^2
      + a_{TT\mA}\big( \lag T \rag_\mR - \lag T \rag_\mS \big) \big( \lag T \rag_\mR \lag \mA \rag_\mR
                                                                   + \lag T \rag_\mS \lag \mA \rag_\mR
                                                                   -2\lag T \rag_\mS \lag \mA \rag_\mS \big) \nn\\
&& ~~~
      + a_{TTTT} \big( \lag \mA \rag_\mR - \lag \mA \rag_\mS \big)^2
                 \big( \lag T \rag_\mR^2 + 2 \lag T \rag_\mR \lag T \rag_\mS - 3 \lag T \rag_\mS^2 \big)
  \big] \ell^8 + O(\ell^{10}),
\eea
\bea
&& S(\r_A , \s_A) = -2 a_{TT}\big( \lag T \rag_\mR - \lag T \rag_\mS \big)^2 \ell^4
                    -3 a_{TTT}\big( \lag T \rag_\mR - \lag T \rag_\mS \big)^2 \big( \lag T \rag_\mR + \lag T \rag_\mS \big) \ell^6 \nn\\
&& ~~~
  -\big[
        2 a_{\mA\mA}\big( \lag \mA \rag_\mR - \lag \mA \rag_\mS \big)^2
      + a_{TT\mA}\big( \lag T \rag_\mR - \lag T \rag_\mS \big) \big( 3 \lag T \rag_\mR \lag \mA \rag_\mR
                                                                   - \lag T \rag_\mR \lag \mA \rag_\mS \nn\\
&& ~~~
                                                                   + \lag T \rag_\mS \lag \mA \rag_\mR
                                                                   -3\lag T \rag_\mS \lag \mA \rag_\mS \big)
      + 4 a_{TTTT} \big( \lag \mA \rag_\mR - \lag \mA \rag_\mS \big)^2
                   \big( \lag T \rag_\mR^2 + \lag T \rag_\mR \lag T \rag_\mS + \lag T \rag_\mS^2 \big)
  \big] \ell^8
\nn\\
&& ~~~
   + O(\ell^{10}).
\eea

\begin{figure}[htbp]
  \centering
  \includegraphics[width=0.4\textwidth]{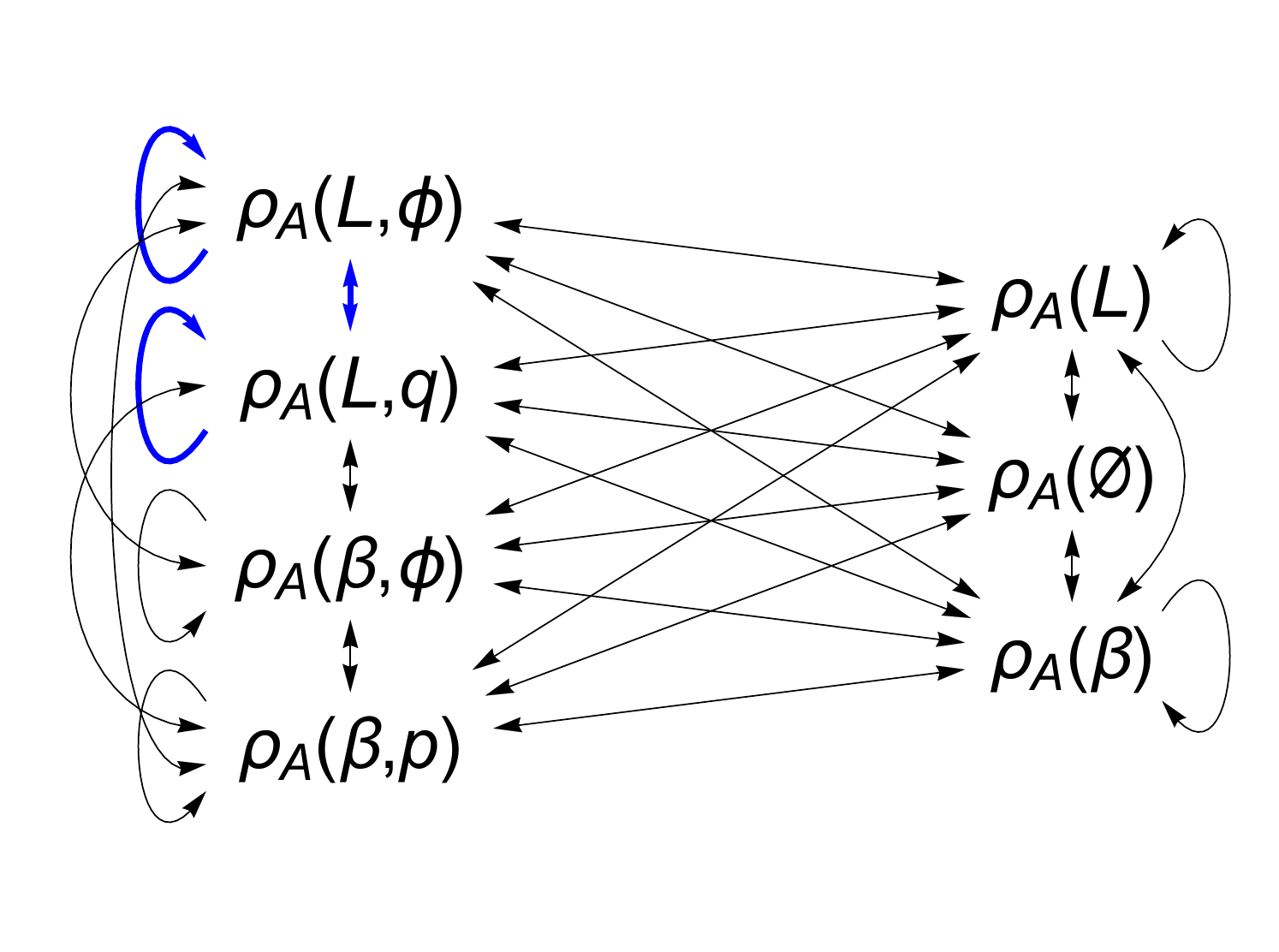}\\
  \caption{The 48 relative entropies we can calculate using OPE of the twist operators. By $\r_A \ra \s_A$ we mean the relative entropy $S(\r_A\|\s_A)$, and by $\r_A \lra \s_A$ we mean the relative entropies $S(\r_A\|\s_A)$ and $S(\s_A\|\r_A)$. Note that $q=\ep^{-2\pi\b/L}$ and $p=\ep^{-2\pi L/\b}$ depend on both $L$ and $\b$. In the figure $\cdots L \cdots \ra \cdots L \cdots$ actually means $\cdots L_1 \cdots \ra \cdots L_2 \cdots$ with generally $L_1 \neq L_2$, $\cdots \phi \cdots \ra \cdots \phi \cdots$ means $\cdots \phi_1 \cdots \ra \cdots \phi_2 \cdots$, and $\cdots \b \cdots \ra \cdots \b \cdots$ means $\cdots \b_1 \cdots \ra \cdots \b_2 \cdots$. In practice, we only need to calculate the four relative entropies as marked in blue.}
  \label{rel}
\end{figure}

As shown in figure~\ref{rel}, we use OPE of twist operators as described in section~\ref{secpre} to calculate four relative entropies. For $\r_A(L_1,\phi_1)$ and $\r_A(L_2,\phi_2)$ we have the relative entropy (\ref{cl1}).
For the special case $L_1=L_2$ in (\ref{cl1}), it matches the result in \cite{He:2017vyf}.
For $\r_A(L_1,\phi)$ and $\r_A(L_2,q)$ we have the relative entropy (\ref{cl2}).
For $\r_A(L_1,q)$ and $\r_A(L_2,\phi)$ we have the relative entropy (\ref{cl3}).
Note that
\be
S(\r_A(L_1,\phi)\|\r_A(L_2,q)) \neq S(\r_A(L_2,q)\|\r_A(L_1,\phi)).
\ee
For $\r_A(L_1,q_1)$ and $\r_A(L_2,q_2)$ we have the relative entropy (\ref{cl4}).

For general $n\neq1$, the $n$-th relative entropy $S_n(\r_A||\s_A)$ and $n$-th symmetrized relative entropy $S_n(\r_A,\s_A)$ have no obvious physical meaning because they are not positive definite. However, the 2nd symmetrized relative entropy, which is defined as
\be
S_2(\r_A,\s_A) = \log \f{(\tr_A\r_A^2)(\tr_A\s_A^2)}{[\tr_A(\r_A\s_A)]^2},
\ee
is positive definite and can be used to characterize the dissimilarity of $\r_A$, $\s_A$. In fact, it is directly related to the overlap of the two reduced density matrices
\be
\mF(\r_A,\s_A) = \f{[\tr_A(\r_A\s_A)]^2}{(\tr_A\r_A^2)(\tr_A\s_A^2)}.
\ee

\begin{figure}[htbp]
  \centering
  \includegraphics[width=0.4\textwidth]{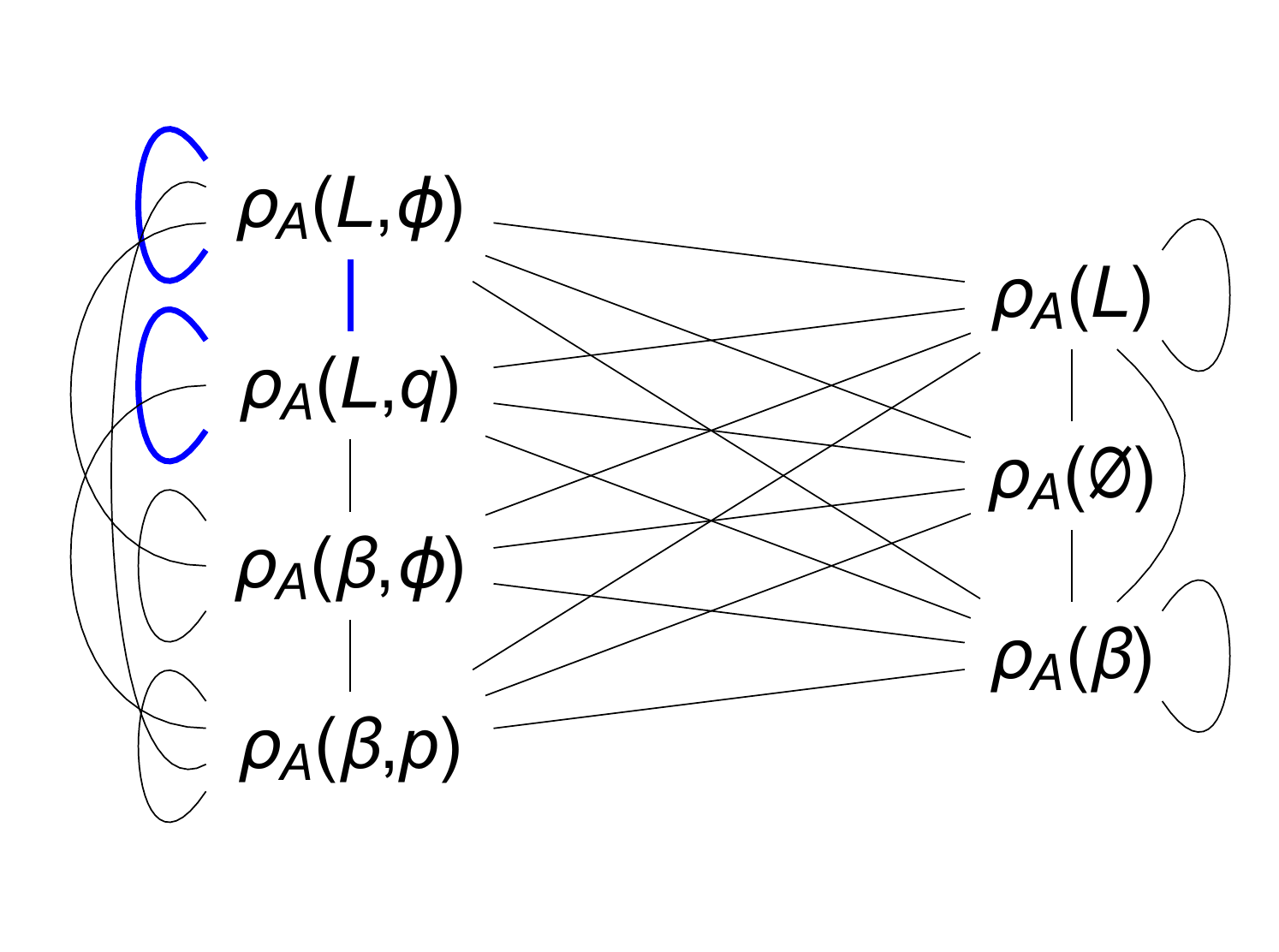}\\
  \caption{The 27 symmetrized relative entropies we can calculate using OPE of the twist operators. We only need to calculate the three ones marked in blue. This figure also applies to the 2nd symmetrized relative entropy, Jensen-Shannon divergence, as well as the Schatten 2-norm and 4-norm in the following subsections.}
  \label{dis}
\end{figure}

As shown in figure~\ref{dis}, we calculate three symmetrized relative entropies (\ref{cl5}), (\ref{cl6}), and (\ref{cl7}) using OPE of twist operators. We get the 2nd symmetrized relative entropies (\ref{cl8}), (\ref{cl9}), and (\ref{cl10}).

\subsection{Jensen-Shannon divergence}\label{secjsd}

The Jensen-Shannon divergence of two reduced density matrices $\r_A$ and $\s_A$ are defined as
\be\label{JEE}
JS(\r_A,\s_A) = S\big(\f{\r_A+\s_A}{2}\big) - \f12 S(\r_A) - \f12 S(\s_A),
\ee
with $S\big(\f{\r_A+\s_A}{2}\big)$, $S(\r_A)$, $S(\s_A)$ being the von Neumann entropies.
By definition
\be
0 \leq JS(\r_A,\s_A) \leq \log 2.
\ee
One can also define the Jensen-Shannon distance
\be
\sqrt{JS(\r_A,\s_A)}.
\ee
To calculate the Jensen-Shannon divergence, we first calculate the Jensen-R\'enyi divergence
\be
JR_n(\r_A,\s_A) = S_n\big(\f{\r_A+\s_A}{2}\big) - \f12 S_n(\r_A) - \f12 S_n(\s_A),
\ee
with $S_n\big(\f{\r_A+\s_A}{2}\big)$, $S_n(\r_A)$, $S_n(\s_A)$ being the R\'enyi entropies, and then take the $n \to 1$ limit.
We then get
\bea
&& JS(\r_A , \s_A) = - \f14 a_{TT}\big( \lag T \rag_\mR - \lag T \rag_\mS \big)^2 \ell^4
                     - \f38 a_{TTT}\big( \lag T \rag_\mR - \lag T \rag_\mS \big)^2 \big( \lag T \rag_\mR + \lag T \rag_\mS \big) \ell^6 \nn\\
&& ~~~
  -\big[
        \f14 a_{\mA\mA}\big( \lag \mA \rag_\mR - \lag \mA \rag_\mS \big)^2
      + \f18 a_{TT\mA}\big( \lag T \rag_\mR - \lag T \rag_\mS \big) \big( 3 \lag T \rag_\mR \lag \mA \rag_\mR
                                                                   - \lag T \rag_\mR \lag \mA \rag_\mS \nn\\
&& ~~~
                                                                   + \lag T \rag_\mS \lag \mA \rag_\mR
                                                                   -3\lag T \rag_\mS \lag \mA \rag_\mS \big)
      + \f1{16} a_{TTTT} \big( \lag \mA \rag_\mR - \lag \mA \rag_\mS \big)^2
                         \big( 7 \lag T \rag_\mR^2 + 10 \lag T \rag_\mR \lag T \rag_\mS + 7 \lag T \rag_\mS^2 \big)
  \big] \ell^8
\nn\\
&& ~~~
   + O(\ell^{10}).
\eea
Explicitly, we obtain (\ref{cl11}), (\ref{cl12}), and (\ref{cl13}).

\subsection{Schatten 2-norm and 4-norm}\label{secs24}

For a general matrix $\r$, the Schatten $n$-norm is defined as
\be
\| \r \|_n = (\tr |\r|^n)^{1/n},
\ee
with $|\r|=\sqrt{\r^\dagger\r}$.
For $n=1$ it is just the trace norm, and for $n=2$ it is just the Hilbert-Schmidt norm.
For two reduced density matrices $\r_A$, $\s_A$, we just calculate
\be \label{y39}
\| \r_A-\s_A \|_n^n = \tr_A |\r_A-\s_A|^n.
\ee
For $n=1$ it is just the trace distance, and for $n=2$ it is just trace square. Since the reduced density matrices are hermitian, when $n:=2p$ is an even integer we have a simpler expression
\be
\| \r_A-\s_A \|_{2p}^{2p} = \tr_A (\r_A-\s_A)^{2p}.
\ee
When there is no ambiguity, we call $\| \r_A-\s_A \|_{2p}^{2p}$ also as Schatten $2p$-norm.
We use (\ref{z32}) and get the Schatten 2-norms (\ref{cl14}), (\ref{cl15}), (\ref{cl16}) and Schatten 4-norms (\ref{cl17}), (\ref{cl18}), (\ref{cl19}).

\section{ETH for canonical ensemble thermal state}\label{secethce}

Whether ETH is satisfied or not depends on how it is precisely defined, and for different quantities there may be different criteria. The local ETH is defined in terms of local operators \cite{Deutsch:1991,Srednicki:1994}.
More precisely it requires that in the basis of energy eigenstates $\{|\phi_a\rag\}$ the operator $A$ has the form
\be
A_{ab} = \bar A(E) \d_{ab} + \ep^{-\mO(S(E))},
\ee
with $E=\f{E_a + E_b}{2}$, $S(E)$ being the microcanonical ensemble entropy with energy $E$, and $\bar A(E)$ being a smooth and slowly varying function of $E$. Under such a condition the expectation value of $A$ with respect to (w.r.t.) a single eigenstate equals the long time average of the expectation value of $A$ w.r.t. a coherent state in a narrow energy window around this single eigenstate, and it also equals to the microcanonical ensemble average of $A$ in this narrow energy window up to exponential suppression of the entropy $S(E)$.
A generalization of local ETH is the subsystem ETH that is defined in terms of reduced density matrices \cite{Lashkari:2016vgj,Dymarsky:2016aqv}, and it states that in the excited state $|\phi\rag$ of energy $E$ the reduced density matrix $\r_{A,\phi}$ of a small region $A$ is close to some universal density matrix $\r_{A,E}$ by trace distance
\be
\| \rho_{A,\phi} - \rho_{A,E} \|_1 \sim \ep^{-\mathcal{O}(S(E))}.
\ee

In this paper we do not check directly the local ETH or subsystem ETH. Instead we compare the reduced density matrix of the excited energy eigenstate with the reduced density matrices of some explicit thermal states. In this section we consider the canonical ensemble states, in section~\ref{secethgge}  the GGE thermal state, and in section~\ref{secme}  the microcanonical ensemble thermal state. We use several quantities to characterize the difference of the reduced density matrices of the excited and thermal states. To claim whether ETH is satisfied or not, we need to set up a criterion for each quantity, which is beyond the scope of the present paper.
Our results can be viewed as a first step towards such criteria.
However, based the observations in \cite{Fitzpatrick:2014vua,Fitzpatrick:2015zha,Asplund:2014coa,Caputa:2014eta,Garrison:2015lva,%
Lashkari:2016vgj,Dymarsky:2016aqv,Lu:2017tbo,Lashkari:2017hwq}, we can make some claims for the R\'enyi entropy and entanglement entropy, as we will discuss in the end of this section.

As a first step towards defining and checking ETH for the canonical ensemble thermal state, we calculate various quantities to characterize the dissimilarity of the reduced density matrix $\r_A(L,\phi)$ for the excited state and $\r_A(\b)$ for the thermal state. Note that  ETH is for comparing a highly exited state and a high temperature state, so that  we use $\r_A(\b)$ to approximate $\r_A(\b,p)$. The excited state $|\phi\rag$ is heavy and we write the conformal weight as
\be
h_\phi=c\e_\phi,
\ee
and by requiring
\be
\lag T \rag_{\mR(L,\phi)} = \lag T \rag_{\mR(\b)},
\ee
we get the identification\cite{Fitzpatrick:2014vua,Fitzpatrick:2015zha}
\be
\b=\f{L}{\sr{24\e_\phi-1}}.
\ee

We have the difference of R\'enyi entropy\footnote{In this equation we omit the order $\ell^8$ part, and denote it by $\cdots$. The full form can found in the attached Mathematica notebook in arXiv. It is the same for other equations with $\cdots$ in this paper.}
\bea \label{DeltaSn}
&& S_n(L,\phi) - S_n(\b) =
~
\frac{\pi^4 c (n-1) (n+1)^2 \epsilon_{\phi} (22 \epsilon_{\phi}-1)\ell^4}{90 n^3 L^4}\\
&& ~~~
~
-\frac{ \pi^6 c (n-1)(n+1)^2 \epsilon_{\phi} [8 (145 n^2+188) \epsilon_{\phi}^2-3 (46 n^2+37) \epsilon_{\phi}+4
   n^2+2]\ell^6}{2835 n^5 L^6}
~
+\cdots\ell^8 + O(\ell^{10}),\nn
\eea
and it has been calculated in \cite{Lashkari:2016vgj,Lin:2016dxa,He:2017vyf}. The difference of entanglement entropy is
\be \label{DeltaS}
S(L,\phi) - S(\b) = -\frac{128 \pi^8 c \epsilon_{\phi}^2(22 \epsilon_{\phi}-1)^2\ell^8}
                          {1575 (5 c+22) L^8}+O(\ell^{10}),
\ee
and it has been calculated in \cite{He:2017vyf}. We have the relative entropies
\bea \label{e68}
&& S(\r_A(L,\phi)\|\r_A(\b)) = \frac{128 \pi^8 c \epsilon_{\phi}^2(22 \epsilon_{\phi}-1)^2\ell^8}
                          {1575 (5 c+22) L^8}+O(\ell^{10}), \nn\\
&& S(\r_A(\b)\|\r_A(L,\phi)) = \frac{128 \pi^8 c \epsilon_{\phi}^2(22 \epsilon_{\phi}-1)^2\ell^8}
                          {1575 (5 c+22) L^8}+O(\ell^{10}),
\eea
and the first one has been calculated in \cite{He:2017vyf} by a different method. Note that $S(\r_A(L,\phi)\|\r_A(\b))$ and $S(\r_A(\b)\|\r_A(L,\phi))$ happen to be the same at order $\ell^8$, and we expect they will be different at higher orders. We have the symmetrized relative entropy and the 2nd symmetrized relative entropy
\bea
&& S(\r_A(L,\phi),\r_A(\b)) = \frac{256 \pi^8 c \epsilon_{\phi}^2(22 \epsilon_{\phi}-1)^2\ell^8}
                          {1575 (5 c+22) L^8}+O(\ell^{10}), \nn\\
&& S_2(\r_A(L,\phi),\r_A(\b)) = \frac{\pi^8 c (5 c+27) \epsilon_{\phi}^2(22 \epsilon_{\phi}-1)^2\ell^8}
                                     {3200 (5 c+22) L^8} + O(\ell^{10}).
\eea
The Jensen-R\'enyi divergence and Jensen-Shannon divergence are respectively
\bea \label{e69}
&& \hspace{-8mm}
   JR_n(\r_A(L,\phi),\r_A(\b)) =
-\frac{ \pi^8(n+1) c \epsilon_{\phi}^2 (22 \epsilon_{\phi}-1)^2\ell^8}
      {2268000 (5 c+22) n^7 L^8 }
[ 175 c^2 (n^2-1)^3+70 c (7 n^2-55) (n^2-1)^2  \nn\\
&& \hspace{-8mm} \phantom{JR_n(\r_A(L,\phi),\r_A(\b)) =}
     -8 (n^2+11) (157 n^4-298 n^2+381) ]
+O(\ell^{10}),\nn\\
&& \hspace{-8mm}
   JS(\r_A(L,\phi),\r_A(\b)) = \frac{32 \pi^8 c \epsilon_{\phi}^2 (22 \epsilon_{\phi}-1)^2\ell^8}
                                    {1575 (5 c+22) L^8} + O(\ell^{10}).
\eea
We also have
\be \label{tsd}
\| \r_A(L,\phi)-\r_A(\b) \|_2^2 = \Big( \f{\ell}{\e} \Big)^{-\f{c}8} \Big[ \frac{\pi^8 c [c (5 c+62)+216]\epsilon_{\phi}^2 (22 \epsilon_{\phi}-1)^2\ell^8}{25600 (5 c+22) L^8}+O(\ell^{10}) \Big].
\ee

As we have said in the beginning of this section, with the above results, we cannot claim whether ETH is satisfied for an individual quantity without a precise criterion of ETH. As stated in \cite{Lashkari:2016vgj,Dymarsky:2016aqv}, in a CFT not every quantity is good to define ETH. For the R\'enyi entropies of the excited and thermal states being equal, it is necessary that the subsystem is much smaller than the whole system $\ell/L \to 0$ \cite{Garrison:2015lva,Lu:2017tbo}. If one defines ETH for canonical ensemble as $S_n(L,\phi) - S_n(\b) \to 0$ when $\ell/L \to 0$, then from (\ref{DeltaSn}) one concludes that such an ETH is satisfied. However, this criterion seems too strong to yield useful result for general cases. Instead, we can think R\'enyi entropy as a refined quantity compared to the entanglement entropy to characterize the violation of local thermality of a energy eigenstate.


Similarly, the Jensen-Shannon divergence is a better quantity to define ETH than the Jensen-R\'enyi divergence, since the former is always nonnegative due to the concavity of the von Neumann entropy while the latter is not. This can be seen in equations (\ref{e69}). Note that at order $\ell^8$, the Jensen-R\'enyi divergence is of order $c^2$ and the Jensen-Shannon divergence of order $c^0$. This is reminiscent of the fact that the R\'enyi entropy difference is of order $c$ and the entanglement entropy difference is of order $c^0$. This is another indication that the Jensen-R\'enyi divergence is not a good quantity to define ETH, as the R\'enyi entropy.
The R\'enyi entropy is just a higher genus free energy, and this is consistent with the fact that it is of order $c$.
However, the Jensen-R\'enyi divergence is not a free energy or a sum of free energies, it is not necessary that it is of order $c$ or subleading to order $c$.

The Schatten 2-norm (\ref{tsd}), or equivalently the square trace distance, is dependent on the UV regulator $\e$ and it is vanishing as $\e/\ell \to 0$. It is not a good quantity to define ETH, either.

For a large $c$ CFT, it was found in \cite{Asplund:2014coa,Caputa:2014eta} that the leading order $c$ entanglement entropy of the excited and canonical ensemble thermal states is the same as long as $0 < \ell/L <1/2$. If ETH for the entanglement entropy is defined in this way with $0 < \ell/L <1/2$, the result (\ref{DeltaS}) clearly shows the violation of ETH at the next-to-leading order of large $c$ \cite{He:2017vyf}.

\section{ETH for GGE thermal state}\label{secethgge}

All the above dissimilarities in the previous section between the excited and thermal state originate from the fact that the level 4 operator $\mA$ has different expectation values \cite{Lin:2016dxa,He:2017vyf}
\be
\lag \mA \rag_{\mR(L,\phi)} \neq \lag \mA \rag_{\mR(\b)}.
\ee
A more refined consideration is that one should not compare the excited state and the canonical ensemble thermal state, instead one need to consider the generalized Gibbs ensemble (GGE) thermal state \cite{Rigol:2006,Mandal:2015jla,Cardy:2015xaa,Mandal:2015kxi,Vidmar:2016,deBoer:2016bov}. The GGE state has the density matrix
\be
\r_\GGE = \ep^{-\b H -\sum_i \b_i J_i},
\ee
with $J_i$ being some conserved charges and $\b_i$ being the corresponding chemical potentials. By requiring the ETH comparison is done for the same macroscopic super-selection sector, we should impose
\be
\lag H \rag_{\mR(L,\phi)} = \lag H \rag_\GGE, ~~
\lag J_i \rag_{\mR(L,\phi)} = \lag J_i \rag_\GGE,
\ee
so that one can get the relation of $h_\phi$ with the GGE parameters $\b$, $\m_i$. In the vacuum conformal family, there are an infinite number of commuting conserved charges $I_{2k+1}$ with $k=0,1,\cdots$ \cite{Sasaki:1987mm,Bazhanov:1994ft}. For examples, one has
\bea
&& I_1 = -\f{1}{2\pi} \int_{-L/2}^{L/2} dw T(w) = \f{2\pi}{L} \Big( L_0 - \f{c}{24} \Big) = H, \\
&& I_3 = \f{1}{2\pi} \int_{-L/2}^{L/2} dw \mA(w) = \Big( \f{2\pi}{L} \Big)^3 \Big[ \mA_0 -\f{5c+22}{60} \big( L_0 - \f{c}{48} \big) \Big]. \nn
\eea

We may choose the GGE state
\be
\r_\GGE = \ep^{-\b H -\sum_{k=1}^\inf \b_{2k+1} I_{2k+1}}.
\ee
Then we have the requirement
\be \label{z44}
\lag \mX \rag_{\mR(L,\phi)} = \lag \mX \rag_\GGE,
\ee
for all vacuum conformal family quasiprimary operator $\mX$. Since there are more equations than the unknown chemical potentials, we do not know if there is a unique solution for all $\b$, $\b_{2k+1}$, $k=1,2,\cdots$. If this is the case, all the dissimilarities considered in this paper vanish so that there is no difference between the reduced density matrices of the excited state and GGE thermal state.

Furthermore, in GGE it is not necessarily that all the conserved charges commute with each other \cite{Cardy:2015xaa}. For each nonidentity quasiprimary operator in vacuum conformal family, say $\mX$, we may define a conserved charge
\be
I_\mX \pp \int_{-L/2}^{L/2} dw \mX(w).
\ee
Then we may define the GGE state
\be
\r_\GGE = \ep^{ -\sum_\mX \b_\mX I_\mX},
\ee
with which there are the same number of equations and the unknown chemical potentials. However, we still do not know if there is any solution to the equations (\ref{z44}).

To be more concrete, we consider a toy model of GGE
\be
\r_\GGE = \ep^{- \b H - \f{\m}{2\pi}\int_{-L/2}^{L/2} d w \mA(w) }.
\ee
For an arbitrary operator $\mX$ we have
\be
\f{\tr(\mX(w_0)\r_\GGE)}{\tr \ep^{-\b H}} =
\Big\lag \mX(w_0) \ep^{ - \f{\m}{2\pi}\int_{-L/2}^{L/2} d w \mA(w) } \Big\rag_{\mR(\b,p)}
\approx \Big\lag \mX(w_0) \ep^{ - \f{\m}{2\pi}\int_{-\inf}^\inf d w \mA(w) } \Big\rag_{\mR(\b)}.
\ee
We get the expectation value of GGE in expansion of the small chemical potential $\m$
\bea
&& \lag \mX(w_0) \rag_\GGE = \f{\tr(\mX(w_0)\r_\GGE)}{\tr \r_\GGE}
\approx \lag \mX(w_0) \rag_{\mR(\b)}
        -\f{\m}{2\pi}\int_{-\inf}^\inf d w
        \big[ \lag \mA(w) \mX(w_0) \rag_{\mR(\b)} \nn\\
&& \phantom{\lag \mX(w_0) \rag_\GGE = \f{\tr(\mX(w_0)\r_\GGE)}{\tr \r_\GGE} \approx}
 - \lag \mA(w)\rag_{\mR(\b)} \lag \mX(w_0) \rag_{\mR(\b)} \big] + O(\m^2).
\eea
The correlation functions on the cylinder $\mR(\b)$ can be calculated by mapping the cylinder to a complex plane by the conformal transformation $z=\ep^{\f{2\pi w}{\b}}$. Note that the above expectation value should be independent of the position $w_0$. Using the integral\footnote{Note that the integral is only convergent for $0<\Re S<1$, and it is analytically continued to other values of $S$. The results are the same as these from more delicate calculations in \cite{Datta:2014ska,Datta:2014uxa,Datta:2014zpa}.}
\be
\int_0^\inf \f{d x}{\sinh^S x} = \f{\G(\f{S}{2})\G(\f{1-S}{2})}{2\sqrt\pi},
\ee
with $S=4$ and $S=8$, we finally get
\bea \label{z51}
&& \lag T \rag_\GGE = -\frac{\pi^2 c}{6 \beta^2}+\frac{\pi^4 c (5 c+22) \mu }{45 \beta^5}+O(\mu^2), \nn\\
&& \lag \mA \rag_\GGE = \frac{\pi^4 c (5 c+22)}{180 \beta^4}-\frac{\pi^6 c (5 c+22) (7 c+74) \mu }{945
   \beta^7}+O(\mu^2).
\eea
In the excited state $|\phi\rag$ of a holomorphic primary operator $\phi$ with conformal weight $h_\phi=c\e_\phi$, there are expectation values \cite{Lin:2016dxa,He:2017vyf}
\be \label{z52}
\lag T \rag_\phi = \frac{\pi^2 c (1 -24 \e_\phi)}{6 L^2}, ~~
\lag\mA \rag_\phi = \frac{\pi^4  c ((5 c+22) -240 (c+2) \e_\phi +2880 c \e_\phi^2 )}{180 L^4}.
\ee

 To consider ETH comparison for the same super-selection sector, we equate (\ref{z51}) and (\ref{z52})
\be \label{z54}
\lag T \rag_\GGE = \lag T \rag_\phi, ~~ \lag \mA \rag_\GGE = \lag \mA \rag_\phi,
\ee
and solve the inverse temperature $\b$ and chemical potential $\m$ in terms of $\e_\phi$, $c$, $L$. As known that the ETH for canonical ensemble works well in the leading order of large $c$ limit \cite{Asplund:2014coa,Caputa:2014eta}, we should then expect
\be \label{z53}
\lim_{c \to \inf} \b = \f{L}{\sqrt{24\e_\phi -1}}, ~~ \lim_{c \to \inf} \m = 0.
\ee
On the other hand, the finite $c$ correction causes the mismatch between excited state and the canonical thermal state by power suppression of $1/c$ \cite{He:2017vyf}, we then need to find the solution of (\ref{z54}) for GGE with power correction of $1/c$ to (\ref{z53}) as follows. To make the $1/c$ expansions in (\ref{z51}) well-defined, we need the leading order $\m \sim 1/c^\a$ with $\a>1$. Since there is no subleading term in $\lag T \rag_\phi$, we need the leading order correction to $\b$ of order $1/c^{\a-1}$. We then make the following ansatz for the solution to equations (\ref{z54})
\bea \label{z55}
&& \b = \f{L}{\sqrt{24\e_\phi -1}} + \f{a L}{c^{\a-1}} + o\Big(\f{1}{c^{\a-1}}\Big), \nn\\
&& \m = \f{b L^3}{c^{\a}} + o\Big(\f{1}{c^{\a}}\Big),
\eea
with the constants $\a$, $a$, $b$ to be determined. It is easy to see that $\lag \mA \rag_\GGE = \lag \mA \rag_\phi$ cannot be satisfied for $\a \geq 2$. Thus, we have $1<\a<2$ in ansatz (\ref{z55}). However, we cannot determine the coefficient $b$ in ansatz (\ref{z55}) at the present expansion order of (\ref{z51}), but might be determined uniquely at the higher expansion orders.\footnote{In a recent paper \cite{Lashkari:2017hwq}, it is argued that ETH for GGE thermal state does not work in perturbation of small chemical potential and one has to calculate the one-point functions non-perturbatively.}



\section{ETH for microcanonical ensemble thermal state}\label{secme}

The local ETH \cite{Deutsch:1991,Srednicki:1994} and its corollaries such as the subsystem ETH \cite{Lashkari:2016vgj,Dymarsky:2016aqv} are originally considered for comparing the energy eigenstate and the microcanonical (ensemble) thermal state. Despite that the difference between canonical and microcanonical thermal states is power-law negligible in the limit of large number of degrees of freedom, it is still interesting to check ETH directly for microcanonical thermal state. In this appendix we will do this using OPE of twist operators as described in section~\ref{secpre}.

   The microcanonical thermal state to be considered is the equal-weight sum of the pure states $|\phi_i\rangle:=\phi_i|0\rangle$ $i=1,2,\cdots,\O$, i.e.,  its density matrix is given by
\be \label{rme}
\r_\me = \f1\O \sum_{i=1}^\O |\phi_i\rag \lag\phi_i|
\ee
where $\phi_i$'s are nonidentity primary operators of conformal weights $(h_{\phi_i}, {\bar h}_{\phi_i})$. For the microcanonical thermal states, we should require for all $i=1,2,\cdots,\O$
\be\label{approximation}
h_{\phi_i}\simeq h_{\phi}, ~~ {\bar h}_{\phi_i} \simeq {\bar h}_{\phi}
\ee
where $(h_{\phi}, {\bar h}_{\phi})$ is the conformal weight of the excited state $\phi$ with which we will compare for checking ETH.

For simplicity, we can choose an orthonormal set of $\phi_i$'s, i.e.,
\be \label{e58}
\lag \phi_i | \phi_j \rag = \d_{ij}.
\ee
We also choose $\phi$ as one of the $\O$ operators $\phi_i$, i.e., $\phi\in\{\phi_i\}$.

Globally, the pure excited state density matrix $\r_\phi = |\phi\rag \lag\phi|$ and the microcanonical thermal state density matrix $\r_\me=\f1\O\sum_{i=1}^\O \r_{\phi_i}$ are very different. This can be seen from various dissimilarity measures, i.e., starting from their von-Neumann entropies,
\be
S(\r_\phi)=0, ~~ S(\r_\me)=\log\O,
\ee
and then the relative entropy
\be
S(\r_\phi\|\r_\me) = \log\O,
\ee
and the Jensen-Shannon divergence
\be
JS(\r_\phi,\r_\me) = \log 2 + \f12 \log\O - \f{\O+1}{2\O}\log(\O+1).
\ee

Instead, the ETH should be explored by the local observables. If ETH holds, for arbitrary local observable  $\mX$ we should have
\be \label{e57}
\tr(\r_\phi\mX) \simeq \tr(\r_\me \mX).
\ee
If $\mX$ is the operator in the vacuum conformal family, it is easy to see that (\ref{e57}) holds by the fact (\ref{approximation}). On the other hand, if $\mX$ is some nonidentity primary operator or its descendants, then the ETH imposes constraints on OPE coefficients $C_{\phi_i\phi_i \mX}$:
\be
C_{\phi\phi\mX} \simeq \f1\O \sum_{i=1}^\O C_{\phi_i\phi_i\mX}.
\ee
This implies that not every CFT satisfies ETH.

   However, in a large $c$ CFT, it is often a good approximation to consider contributions only from the vacuum conformal family, and this is what we adopt in this paper. We now consider to divide the circle of length $L$, on which the large $c$ CFT lives, into a small subsystem $A$ of length $\ell$ and its large compliment $B$ of length $L-\ell$.  We can define the reduced density matrices $\r_{A,\phi}$ and $\r_{B,\phi}$ for the excited state $\r_\phi$, and $\r_{A,\me}$ and $\r_{B,\me}$ for the microcanonical thermal state $\r_\me$. We then use OPE of twist operators to calculate dissimilarity measures for  comparing $\r_{A,\phi}$, $\r_{A,\me}$, and  for comparing $\r_{B,\phi}$, $\r_{B,\me}$.

   We only include contributions from the vacuum conformal family in the following calculation. For the small subsystem $A$, from (\ref{e57}) we get
\be
\tr_A(\r_{A,\phi}^m\r_{A,\me}^{n-m}) \simeq \tr_A\r_{A,\phi}^n, ~~ m=0,1,\cdots,n,
\ee
and we further get the entanglement entropy, relative entropy, and Jensen-Shannon divergence
\be \label{e60}
S(\r_{A,\phi}) \simeq S(\r_{A,\me}), ~~
S(\r_{A,\phi}\|\r_{A,\me})\simeq 0, ~~
JS(\r_{A,\phi},\r_{A,\me})\simeq 0.
\ee
For the large subsystem $B$, we use (\ref{e58}) and \cite{Chen:2014ehg}
\be
\tr_B[\r_B(\phi_{i_1},\phi_{i_1})\r_B(\phi_{i_2},\phi_{i_2})\cdots\r_B(\phi_{i_n},\phi_{i_n})]
=\tr_A[\r_A(\phi_{i_1},\phi_{i_2})\r_A(\phi_{i_2},\phi_{i_3})\cdots\r_A(\phi_{i_n},\phi_{i_1})],
\ee
and get
\be
\tr_B\r_{B,\me}^n \simeq \O^{1-n}\tr_A\r_{A,\phi}^n, ~~
\tr_B( \r_{B,\phi}^m\r_{B,\me}^{n-m} ) \simeq \O^{m-n}\tr_A\r_{A,\phi}^n, ~~ m=1,2,\cdots,n.
\ee
Then we get
\bea \label{e61}
&& S(\r_{B,\phi}) = S(\r_{A,\phi}), ~~
   S(\r_{B,\me}) \simeq S(\r_{A,\phi}) + \log\O, \nn\\
&& S(\r_{B,\phi}\|\r_{B,\me}) \simeq \log\O, \nn\\
&& JS(\r_{B,\phi},\r_{B,\me})\simeq \log 2 + \f12 \log\O - \f{\O+1}{2\O}\log(\O+1).
\eea

The above result agrees with the expectation from ETH, which states that the energy eigenstate approximates the microcanonical thermal state only for a small enough subsystem but not for a large one. This is also verified by the numerical simulations for lattice models done in \cite{Garrison:2015lva} as long as the size of subsystem is smaller than the half of the total system size. When the size of the subsystem $A$ becomes as large as half the total system size, the trace square distance starts to deviate from zero, and the behavior indicates that one may be able to extract some critical exponents from the behavior around $\ell=L/2$.

As a byproduct, from (\ref{e60}), (\ref{e61}), we get the approximate saturation of the microcanonical ensemble version of the Araki-Lieb inequality \cite{Araki:1970ba}
\be \label{e62}
S(\r_{\me}) - S(\r_{B,\me}) + S(\r_{A,\me}) \simeq 0.
\ee
The saturation of canonical ensemble version of the Araki-Lieb inequality and its holography have been studies in \cite{Ryu:2006bv,Headrick:2007km,Azeyanagi:2007bj,Blanco:2013joa,Hubeny:2013gta,Faulkner:2013ana,Chen:2014ehg,Chen:2017ahf}.
We expect the suturation in (\ref{e62}) is lifted if the approximation (\ref{approximation}) is scrutinized carefully and/or contributions from nonvacuum conformal families are included, i.e., that
\be
S(\r_{\me}) - S(\r_{B,\me}) + S(\r_{A,\me}) > 0.
\ee

\section{Conclusion and discussion}\label{seccon}

We have used the OPE of the twist operators to calculate various quantities that can be used to characterize the dissimilarity of two reduced density matrices, and these quantities include the R\'enyi entropy, entanglement entropy, relative entropy, Jensen-Shannon divergence, as well as the Schatten 2-norm and 4-norm. We first consider contributions from only the holomorphic sector of the vacuum conformal family, and make expansion of all the quantities by the length of short interval $\ell$ to order $\ell^9$.
As an application of the results, for ETH we show how these dissimilarity measures behave for the excited and thermal states in the high temperature limit.
As we have showed in this paper, all these quantities can capture the dissimilarity of the two reduced density matrices. Furthermore, we also discuss the possibility to define ETH with GGE thermal state. 
By using GGE, we provide a possible scenario to define ETH and resolve the mismatch between ETH and highly excited states in large $c$ CFT.
We also discuss ETH for microcanonical ensemble thermal state.
In the appendix we give the formal forms of the entanglement entropy, relative entropy, Jensen-Shannon divergence, and Fisher quantum metric with contributions from general operators.

In the method of twist operators we cannot calculate the trace distance, which is essential for the definition of subsystem ETH \cite{Lashkari:2016vgj,Dymarsky:2016aqv}. Trace distance is just the Schatten $n$-norm with $n=1$, and the absolute value in the definition makes it hard to evaluate when $n$ is an odd integer. It would be nice if the trace distance can be calculated in CFT.



\section*{Acknowledgement}

We would like to thank Alexandre Belin, Xi Dong, Thomas Faulkner, Nabil Iqbal, Zuhair U.\ Khandker, Guojing Liu, G\'abor S\'arosi and Huajia Wang for helpful discussions.
We thank the anonymous JHEP referee of our previous paper \cite{He:2017vyf} for discussions about higher order conserved charges the generalized Gibbs ensemble.
SH is supported by Max-Planck fellowship in Germany and the National Natural Science Foundation of China Grant No.~11305235.
FLL is supported by Taiwan Ministry of Science and Technology through Grant No.~103-2112-M-003-001-MY3, No.~106-2112-M-003-004-MY3 and No.~103-2811-M-003-024.
JJZ is supported by the ERC Starting Grant 637844-HBQFTNCER and in part by Italian Ministero dell'Istruzione, Universit\`a e Ricerca (MIUR) and Istituto Nazionale di Fisica Nucleare (INFN) through the ``Gauge Theories, Strings, Supergravity'' (GSS) research project.

\appendix

\section{Relative entropy from modular Hamiltonian}\label{apprel}

\begin{figure}[htbp]
  \centering
  \includegraphics[height=0.3\textwidth]{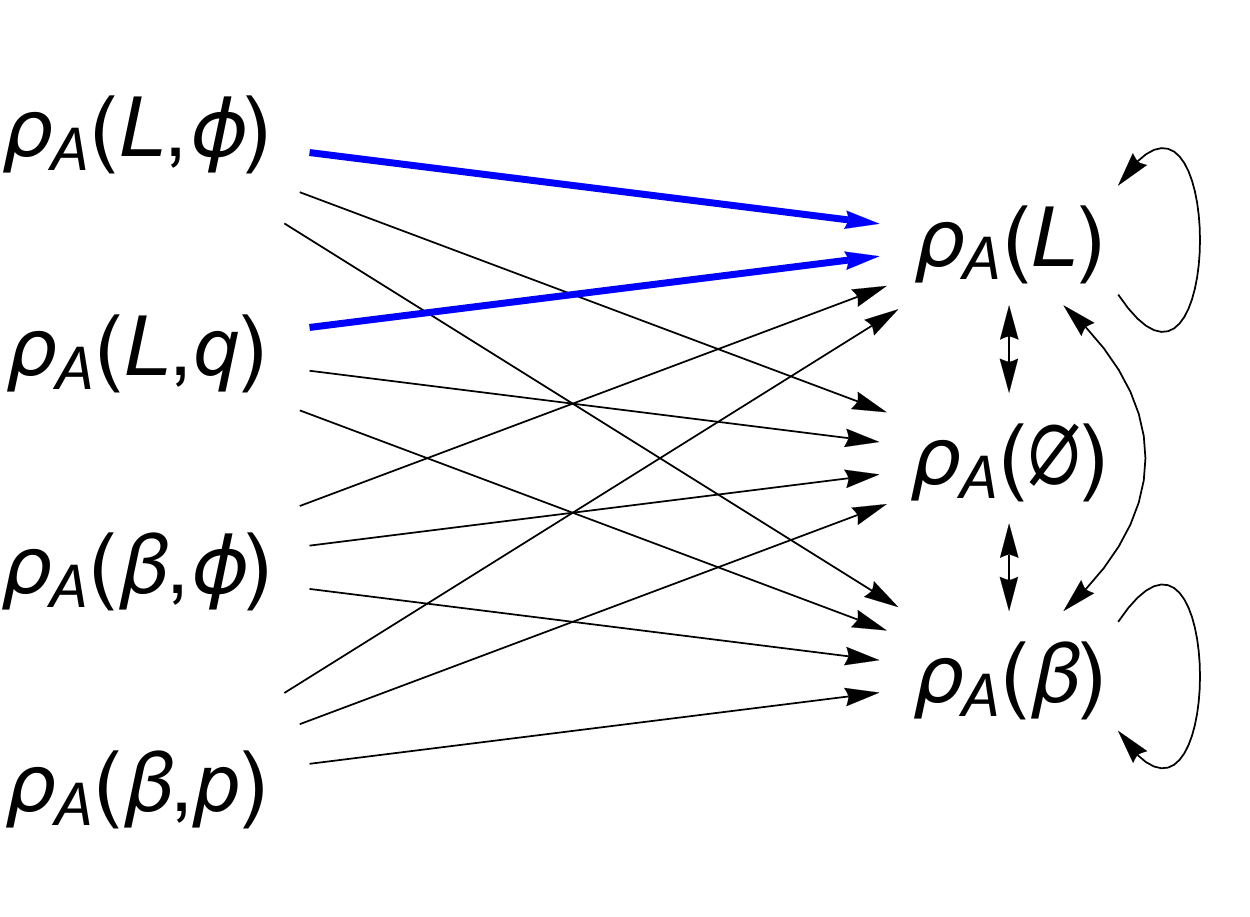}\\
  \caption{The 20 relative entropies we can calculate using modular Hamiltonian and entanglement entropy.  We only need to calculate the two relative entropies $S(\r_A(L_1,\phi)\|\r_A(L_2))$ and $S(\r_A(L_1,q)\|\r_A(L_2))$ as marked in blue.}
  \label{relmod}
\end{figure}

We calculate the relative entropies using modular Hamiltonian as shown in figure~\ref{relmod}, and some of them have been calculated from the same method in \cite{Sarosi:2016oks,Sarosi:2016atx}. This appendix serves as a check of the relative entropies from twist operators in section~\ref{secrel}.

For a reduced density matrix $\r_A$, the modular Hamiltonian $H(\r_A)$ is defined as
\be
\r_A = \f{\ep^{-H(\r_A)}}{\tr_A\ep^{-H(\r_A)}}.
\ee
For two reduced density matrices $\r_A$, $\s_A$, the relative entropy can be written as
\be
S(\r_A\|\s_A) = \lag H(\s_A) \rag_{\r} - \lag H(\s_A) \rag_{\s} -S(\r_A) +S(\s_A),
\ee
with $H(\s_A)$ being the modular Hamiltonian of $\s_A$.
The modular Hamiltonian is known only for cases of $\r_A(\es)$, $\r_A(L)$ and $\r_A(\b)$, and one has \cite{Casini:2011kv,Wong:2013gua,Cardy:2016fqc}\footnote{One can see modular Hamiltonian for excited states in \cite{Lashkari:2015dia,Sarosi:2017rsq}.}
\bea
&& H_A(\es)  = - \int_{0}^{\ell} \f{x(\ell-x)}{\ell}T(x)dx, \nn\\
&& H_{A}(L) = - \f{L}{\pi} \int_{0}^{\ell} \f{\sin\f{\pi x}{L}\sin\f{\pi(\ell-x)}{L}}
                                                     {\sin\f{\pi\ell}{L}}
                                                   T(x)dx, \nn\\
&& H_{A}(\b) = - \f{\b}{\pi} \int_{0}^{\ell} \f{\sinh\f{\pi x}{\b}\sinh\f{\pi(\ell-x)}{\b}}
                                                       {\sinh\f{\pi\ell}{\b}}
                                                     T(x)dx.
\eea
We have only incorporated contributions from the holomorphic sector.
They satisfy the relations
\be
H_A(\es) = \lim_{L\to\inf} H_{A}(L) = \lim_{\b\to\inf} H_{A}(\b), ~~
H_{A}(\b) = H_{A}(L)|_{L\to\ii\b}.
\ee
As shown in figure~\ref{relmod}, we use the entanglement entropy and modular Hamiltonian to calculate the relative entropies $S(\r_A(L_1,\phi)\|\r_A(L_2))$ and $S(\r_A(L_1,q)\|\r_A(L_2))$. We have
\bea
&& S(\r_A(L_1,\phi)\|\r_A(L_2)) =
\frac{\pi^4[c L_1^2-L_2^2 (c-24 h_{\phi})]^2 \ell^4 }{1080 c L_1^4 L_2^4}\\
&& ~~~
+\frac{\pi^6 [c L_1^2-L_2^2 (c-24 h_{\phi})]^2 [2 c L_1^2 +L_2^2 (c-24 h_{\phi})]\ell^6}{17010 c^2 L_1^6 L_2^6}
+ \cdots \ell^8
+O(\ell^{10}), \nn
\eea
and this is in accord with (\ref{cl1}) and (\ref{cl2}).
We have
\bea
&& S(\r_A(L_1,q)\|\r_A(L_2)) =
\Big[ \frac{\pi^4 c (L_1^2-L_2^2)^2}{1080 L_1^4 L_2^4}
     +\frac{4 \pi^4 (L_1^2-L_2^2) q^2}{45 L_1^4 L_2^2}
     +\frac{2 \pi^4 (L_1^2-L_2^2) q^3}{15 L_1^4 L_2^2} \nn\\
&& ~~~
     +\frac{4 \pi^4 [c L_1^2-(c-8) L_2^2] q^4}{15 c L_1^4 L_2^2}
     +O(q^5) \Big]\ell^4
+\Big[ \frac{\pi^6 c (2 L_1^6-3 L_2^2 L_1^4+L_2^6)}{17010 L_1^6 L_2^6}
      +\frac{8 \pi^6 (L_1^4-L_2^4) q^2}{945 L_1^6 L_2^4} \nn\\
&& ~~~
      +\frac{4 \pi^6 (L_1^4-L_2^4) q^3}{315 L_1^6 L_2^4}
      +\frac{8 \pi^6 [c L_1^4-(c-16) L_2^4] q^4}{315 c L_1^6 L_2^4}
      +O(q^5) \Big]\ell^6
+\Big[ \frac{\pi^8 c (3 L_1^8-4 L_2^2 L_1^6+L_2^8)}{226800 L_1^8 L_2^8}\\
&& ~~~
      +\frac{4 \pi^8 (L_1^6-L_2^6) q^2}{4725 L_1^8 L_2^6}
      +\frac{2 \pi^8 (L_1^6-L_2^6) q^3}{1575 L_1^8 L_2^6}
      +\frac{4 \pi^8 [c L_1^6+(159 c+728) L_2^6] q^4}{1575 c L_1^8L_2^6}
      +O(q^5) \Big]\ell^8
+O(\ell^{10}), \nn
\eea
and this is in accord with (\ref{cl3}) and (\ref{cl4}).

\section{Contributions from general operators}\label{secgen}

In the main text, we only consider the contributions from the holomorphic part of the vacuum conformal family to order $\ell^9$. In this appendix we consider the contributions from general holomorphic and antiholomorphic operators, and we get closed forms of the entanglement entropy, relative entropy, and Jensen-Shannon divergence.

For a short interval $A=[0,\ell]$ on a Riemann surface $\mR$ that has translational symmetry, we have the reduced density matrix $\r_A$ and get
\be
\tr_A \r_A^n = \Big( \f{\e}{\ell} \Big)^{4h_\s} \Big( 1 + \sum_{k=1}^n \sum_{\{\mX_1, \cdots, \mX_k\}} \ell^{\D_{\mX_1} + \cdots + \D_{\mX_k}} b_{\mX_1\cdots\mX_k} \lag\mX_1\rag_\mR \cdots \lag\mX_k\rag_\mR \Big),
\ee
with the summation $\{\mX_1\cdots\mX_k\}$ being over different sets of all the nonidentity holomorphic and antiholomorphic quasiprimary operators. For a quasiprimary operator $\mX$, we use $\D_\mX$ to denote its scaling dimension. Then we get the entanglement entropy
\be
S(\r_A) = \f{c}{3} \log\f{\ell}{\e} + \sum_{k=1}^\inf \sum_{\{\mX_1, \cdots, \mX_k\}} \ell^{\D_{\mX_1} + \cdots + \D_{\mX_k}} a_{\mX_1\cdots\mX_k} \lag\mX_1\rag_\mR \cdots \lag\mX_k\rag_\mR.
\ee
For the same short interval $A=[0,\ell]$ on another Riemann surface $\mS$ that also has translation symmetry, we have the reduced density matrix $\s_A$ and similar expression for entanglement entropy $S(\s_A)$. The difference of entanglement entropies is
\be \label{j1}
S(\r_A) - S(\s_A) = \sum_{k=1}^\inf \sum_{\{\mX_1, \cdots, \mX_k\}} \ell^{\D_{\mX_1} + \cdots + \D_{\mX_k}} a_{\mX_1\cdots\mX_k} ( \lag\mX_1\rag_\mR \cdots \lag\mX_k\rag_\mR - \lag\mX_1\rag_\mS \cdots \lag\mX_k\rag_\mS ).
\ee

For $k$ quasiprimary operators $\mX_1, \cdots, \mX_k$, and two translation invariant Riemann surfaces $\mR$, $\mS$, we may define
\be
\mF_i(\mX_1, \cdots, \mX_k|\mR,\mS) = \f{1}{k!} ( \lag\mX_1\rag_\mR \cdots \lag\mX_i\rag_\mR
                                                  \lag\mX_{i+1}\rag_\mS \cdots \lag\mX_k\rag_\mS + \textrm{permutations} ),
\ee
with $0\leq i \leq k$. Note that the above definition is normalized such that $\mF_i(\mX, \cdots, \mX|\mR,\mS) = \lag\mX\rag_\mR^i \lag\mX\rag_\mS^{k-i}$, $\mF_i(\mX_1, \cdots, \mX_k|\mR,\mR) = \lag\mX_1\rag_\mR \cdots \lag\mX_k\rag_\mR$. For $0 \leq m \leq n$, we have
\be \label{z45}
\tr_A ( \r_A^m \s_A^{n-m} ) = \Big( \f{\e}{\ell} \Big)^{4h_\s}
\Big( 1 + \sum_{k=1}^n \sum_{\{\mX_1, \cdots, \mX_k\}} \ell^{\D_{\mX_1} + \cdots + \D_{\mX_k}} b_{\mX_1\cdots\mX_k}
      \sum_{i=0}^k \f{C_m^i C_{n-m}^{k-i}}{C_n^k} \mF_i(\mX_1,\cdots,\mX_k|\mR,\mS) \Big).
\ee
Then we get the relative entropy
\bea \label{j2}
&& S(\r_A\|\s_A) = \sum_{k=2}^\inf \sum_{\{\mX_1, \cdots, \mX_k\}} \big\{ \ell^{\D_{\mX_1} + \cdots + \D_{\mX_k}} a_{\mX_1\cdots\mX_k} [
k \mF_1(\mX_1,\cdots,\mX_k|\mR,\mS)
- \lag\mX_1\rag_\mR \cdots \lag\mX_k\rag_\mR \nn\\
&& \phantom{S(\r_A\|\s_A) =}
- (k-1) \lag\mX_1\rag_\mS \cdots \lag\mX_k\rag_\mS ] \big\},
\eea
and the symmetrized relative entropy
\bea
&& S(\r_A,\s_A) = \sum_{k=2}^\inf k \sum_{\{\mX_1, \cdots, \mX_k\}} \big\{ \ell^{\D_{\mX_1} + \cdots + \D_{\mX_k}} a_{\mX_1\cdots\mX_k} [
 \mF_1(\mX_1,\cdots,\mX_k|\mR,\mS) + \mF_1(\mX_1,\cdots,\mX_k|\mS,\mR) \nn\\
&& \phantom{S(\r_A,\s_A) =}
- \lag\mX_1\rag_\mR \cdots \lag\mX_k\rag_\mR
- \lag\mX_1\rag_\mS \cdots \lag\mX_k\rag_\mS ] \big\}.
\eea

Using the summation
\be
\sum_{m=0}^n C_n^m C_m^i C_{n-m}^{k-i} = 2^{n-k}C_n^k C_k^i,
\ee
and (\ref{z45}), we get
\be
\tr_A \big( \f{\r_A+\s_A}{2}\big)^n = \Big( \f{\e}{\ell} \Big)^{4h_\s} \Big( 1 + \sum_{k=1}^n \sum_{\{\mX_1, \cdots, \mX_k\}} \ell^{\D_{\mX_1} + \cdots + \D_{\mX_k}} b_{\mX_1\cdots\mX_k}
\sum_{i=0}^k \f{C_k^i}{2^k} \mF_i(\mX_1,\cdots,\mX_k|\mR,\mS)  \Big).
\ee
Then we get the Jensen-Shannon divergence
\bea \label{j3}
&& JS(\r_A,\s_A) = \sum_{k=2}^\inf \sum_{\{\mX_1, \cdots, \mX_k\}} \Big\{ \ell^{\D_{\mX_1} + \cdots + \D_{\mX_k}} a_{\mX_1\cdots\mX_k}
\Big[
 - \f12 \lag\mX_1\rag_\mR \cdots \lag\mX_k\rag_\mR
 - \f12 \lag\mX_1\rag_\mS \cdots \lag\mX_k\rag_\mS \nn\\
&& \phantom{S(\r_A,\s_A) =}
 + \sum_{i=0}^k \f{C_k^i}{2^k} \mF_i(\mX_1,\cdots,\mX_k|\mR,\mS)
\Big] \Big\}.
\eea

With the above results, we can also calculate short interval expansion of the Fisher information metric. We parameterize the states of the CFT by $\th^\a$, and we have the density matrix $\r(\th)$, and formally the Riemann surface $\mR(\th)$. For the reduced density matrix $\r_A(\th)$, the Fisher information metric is defined as
\be
g_{\a\b} = \f12 \lt\{ \tr_A [ \r_A(\th) ( \p_\a \log \r_A(\th) ) ( \p_\b \log \r_A(\th) ) ] + (\a\lra\b) \rt\}.
\ee
It is related to the relative entropy and Jensen-Shannon divergence as \cite{Sanchez-Moreno:2010,Crooks:2007}
\bea
&& S(\r_A(\th+\d\th)\|\r_A(\th)) = \f12 g_{\a\b}\d\th^\a\d\th^\b + O[(\d\th)^3], \nn\\
&& JS(\r_A(\th+\d\th),\r_A(\th)) = \f18 g_{\a\b}\d\th^\a\d\th^\b + O[(\d\th)^3].
\eea
From  (\ref{j2}) or (\ref{j3}) we get short interval expansion of the Fisher information metric
\be
g_{\a\b} = - \sum_{k=2}^\inf \sum_{\{\mX_1, \cdots, \mX_k\}} \ell^{\D_{\mX_1} + \cdots + \D_{\mX_k}} a_{\mX_1\cdots\mX_k}
           \mG_{\a\b}(\mX_1,\cdots,\mX_k|\mR(\th)),
\ee
with the definition
\be
\mG_{\a\b}(\mX_1,\cdots,\mX_k|\mR(\th)) = \sum_{1\leq i < j \leq k}
\{ \lag\mX_1\rag_{\mR(\th)} \dots
[\p_{\th^\a}\lag\mX_{i}\rag_{\mR(\th)}]\dots
[\p_{\th^\b}\lag\mX_{j}\rag_{\mR(\th)}]\dots
\lag\mX_k\rag_{\mR(\th)}
+ (\a \lra \b) \}.
\ee
In principle, the Fisher information metric can be used to define the distance on the state space, i.e., all the thermal and quasi-primary states of 2D CFTs as considered in this paper. Though we do not know at present how to efficiently characterize the state space by this metric, we expect it may help to visualize the ETH geometrically for the future studies.

In section~\ref{secethce}, for the reduced density matrices of the excited state and canonical ensemble thermal state $\r_A(L,\phi)$, $\r_A(\b)$, we have calculated the relative entropy $S(\r_A(L,\phi)\|\r_A(\b))$ (\ref{e68}) and Jensen-Shannon divergence $JS(\r_A(L,\phi), \r_A(\b))$ (\ref{e69}), with contributions of only the vacuum conformal family, and find that they are non-vanishing and positive at order $\ell^8$, $c^0$.
One question is can they be cancelled with the addition of some suitable non-vacuum conformal families. We address the issue below.

For a general fermionic operator $\psi$, we have
\be
\lag \psi \rag_{\r(L,\phi)} = \lag \psi \rag_{\r(\b)} = 0.
\ee
Without loss of generality, we consider a hermitian nonidentity bosonic primary operator $\mX$ with normalization $\a_\mX$, scaling dimension $\D_\mX = h_\mX + \bar h_\mX$ and spin $s_\mX= h_\mX - \bar h_\mX$.
Note that $s_\mX$ is an integer, and so $\ii^{4s_\mX}=1$.
From (\ref{j2}), we get the leading correction of conformal family $\mX$ to the relative entropy $S(\r_A(L,\phi)\|\r_A(\b))$
\be
\d_\mX S(\r_A(L,\phi)\|\r_A(\b)) = -\ell^{2\D_\mX} a_{\mX\mX} [ \lag \mX \rag_{\r(L,\phi)} - \lag \mX \rag_{\r(\b)} ]^2 + O(\ell^{2\D_\mX+1},\ell^{3\D_\mX}).
\ee
It turns out that \cite{Chen:2017ahf}
\be
a_{\mX\mX} = - \f{\ii^{2s_\mX}\sqrt{\pi}\G(\D_\mX+1)}{2^{2(\D_\mX+1)}\a_\mX\G(\D_\mX+3/2)}, ~~
\lag \mX \rag_{\r(L,\phi)} = \Big(\f{2\pi}{L}\Big)^{\D_\mX}\ii^{s_\mX}C_{\phi\mX\phi}, ~~
\lag \mX \rag_{\r(\b)} =0,
\ee
and we get
\be \label{e70}
\d_\mX S(\r_A(L,\phi)\|\r_A(\b)) =
\f{\sqrt{\pi}\G(\D_\mX+1)}{4\G(\D_\mX+3/2)} \Big( \f{\pi\ell}{L} \Big)^{2\D_\mX}\f{C_{\phi\mX\phi}^2}{\a_\mX}
+ O(\ell^{2\D_\mX+1},\ell^{3\D_\mX}).
\ee

Since $\mX$ is hermitian, $\a_\mX$ is real and positive $\a_\mX>0$, and on a complex plane we have \cite{DiFrancesco:1997nk,Blumenhagen:2009zz}
\be
[\mX(z,\bar z)]^\dagger = \bar z^{-2 h_\mX} z^{-2 \bar h_\mX} \mX(1/\bar z,1/z),
\ee
and by definition $\phi$ is also a hermitian primary operator $[\phi(0)]^\dagger = \phi(\inf)$. From the three-point correlation function on complex plane
\be
\lag \phi(\inf)\mX(z,\bar z)\phi(0) \rag_\mC = \f{C_{\phi\mX\phi}}{z^{h_\mX}\bar z^{\bar h_\mX}},
\ee
we get that $C_{\phi\mX\phi}$ is real. When $C_{\phi\mX\phi}=0$, the conformal family $\mX$ does not contribute to $S(\r_A(L,\phi)\|\r_A(\b))$, and so we only need to consider the case that $C_{\phi\mX\phi}$ is real and non-vanishing. For (\ref{e70}), we have
\be
\d_\mX S(\r_A(L,\phi)\|\r_A(\b)) > 0.
\ee

Similarly, from (\ref{j3}), we get that the leading correction of the conformal family $\mX$ to the Jensen-Shannon divergence $JS(\r_A(L,\phi), \r_A(\b))$ (\ref{e69}) is real and positive
\be
\d_\mX JS(\r_A(L,\phi),\r_A(\b)) =
\f{\sqrt{\pi}\G(\D_\mX+1)}{16\G(\D_\mX+3/2)} \Big( \f{\pi\ell}{L} \Big)^{2\D_\mX}\f{C_{\phi\mX\phi}^2}{\a_\mX}
+ O(\ell^{2\D_\mX+1},\ell^{3\D_\mX}) > 0.
\ee

In summary, in a unitary CFT, the non-vanishing results of the relative entropy $S(\r_A(L,\phi)\|\r_A(\b))$ (\ref{e68}) and Jensen-Shannon divergence $JS(\r_A(L,\phi), \r_A(\b))$ (\ref{e69}) with contributions of only the vacuum conformal family cannot be cancelled by the addition of any non-vacuum conformal families.

\section{Collection of results in section~\ref{secmeasure}}\label{appcl}

In this appendix we collect some lengthy equations in section~\ref{secmeasure}. In these equations we also omit some complex parts, and denote them by $\cdots$. The full forms can found in the attached Mathematica notebook in arXiv.

\underline{Relative entropy}
\bea \label{cl1}
&& S(\r_A(L_1,\phi_1)\|\r_A(L_2,\phi_2)) =
\frac{\pi^4 [L_1^2 (c-24 h_{\phi_2})-L_2^2 (c-24 h_{\phi_1})]^2\ell^4}{1080 c L_1^4 L_2^4} \nn\\
&& ~~~
+\frac{\pi^6[L_1^2 (c-24 h_{\phi_2})-L_2^2 (c-24 h_{\phi_1})]^2 [2 L_1^2 (c-24 h_{\phi_2})+L_2^2 (c-24 h_{\phi_1})]\ell^6}{17010 c^2 L_1^6 L_2^6} \nn\\
&& ~~~
   +\cdots \ell^8 + O(\ell^{10}),
\eea
\bea \label{cl2}
&&S(\r_A(L_1,\phi)\|\r_A(L_2,q)) =
\Big[ \frac{\pi^4 [c L_1^2- (c-24 h_{\phi} ) L_2^2]^2}{1080 c L_1^4 L_2^4}
     +\frac{4 \pi^4 [ (c-24 h_{\phi} ) L_2^2-c L_1^2] q^2}{45 c L_1^2 L_2^4} \nn\\
&& ~~~
     +\frac{2 \pi^4 [ (c-24 h_{\phi} ) L_2^2-c L_1^2] q^3}{15 c L_1^2 L_2^4}
     +\frac{4 \pi^4 [ (c-24 h_{\phi} ) L_2^2-(c-8) L_1^2] q^4}{15 c L_1^2 L_2^4} +O (q^5) \Big]\ell^4 \nn\\
&& ~~~
+\Big[ \frac{\pi^6 [c L_1^2-(c-24 h_{\phi}) L_2^2]^2 (2 c L_1^2+(c-24 h_{\phi}) L_2^2)}{17010 c^2 L_1^6 L_2^6}
      +\frac{16 \pi^6 [(c-24 h_{\phi}) L_2^2-c L_1^2] q^2}{945 c L_1^2 L_2^6} \nn\\
&& ~~~
      +\frac{8\pi^6 [(c-24 h_{\phi}) L_2^2-c L_1^2] q^3}{315 c L_1^2 L_2^6}
      +\frac{16 \pi^6 [(c-8) (c-24 h_{\phi}) L_2^2-c (c-16) L_1^2] q^4}{315 c^2 L_1^2 L_2^6}
      +O(q^5) \Big]\ell^6 \nn\\
&& ~~~
+\cdots \ell^8
+O(\ell^{10}),
\eea
\bea \label{cl3}
&& S(\r_A(L_1,q)\|\r_A(L_2,\phi)) =
        \Big[ \frac{\pi^4 [(c-24 h_{\phi} ) L_1^2-c L_2^2]^2}{1080 c L_1^4 L_2^4}
             +\frac{4 \pi^4 [(c-24h_{\phi} ) L_1^2-c L_2^2] q^2}{45 c L_1^4 L_2^2} \nn\\
&& ~~~
             +\frac{2 \pi^4 [(c-24 h_{\phi} ) L_1^2-c L_2^2] q^3}{15 c L_1^4 L_2^2}
             +\frac{4 \pi^4 [(c-24 h_{\phi} ) L_1^2-(c-8) L_2^2] q^4}{15 c L_1^4 L_2^2}
             +O (q^5 ) \Big]\ell^4  \nn\\
&& ~~~
      + \Big[ \frac{\pi^6  [ (c-24 h_{\phi} ) L_1^2-c L_2^2]^2 [2 (c-24 h_{\phi}) L_1^2+c L_2^2]}{17010 c^2 L_1^6 L_2^6}
             +\frac{8 \pi^6 [(c-24 h_{\phi})^2 L_1^4-c^2 L_2^4] q^2}{945 c^2 L_1^6L_2^4}\nn\\
&& ~~~
             +\frac{4 \pi^6 [(c-24 h_{\phi})^2 L_1^4-c^2 L_2^4] q^3}{315 c^2 L_1^6 L_2^4}
             +\frac{8 \pi^6 [(c-24 h_{\phi})^2 L_1^4-(c-16) c L_2^4] q^4}{315 c^2 L_1^6 L_2^4}+O(q^5) \Big]\ell^6 \nn\\
&& ~~~
+\cdots \ell^8
+O(\ell^{10}),
\eea
\bea \label{cl4}
&& S(\r_A(L_1,q_1)\|\r_A(L_2,q_2)) =
\Big[ \frac{\pi^4 c (L_1^2-L_2^2)^2}{1080 L_1^4 L_2^4}
     -\frac{4 \pi^4 (L_1^2-L_2^2) (L_1^2q_2^2-L_2^2 q_1^2)}{45 (L_1^4 L_2^4)} \nn\\
&& ~~~
     -\frac{2 \pi^4 (L_1^2-L_2^2) (L_1^2 q_2^3-L_2^2q_1^3)}{15 (L_1^4 L_2^4)}
    -\frac{4 \pi^4 [(c-8)(L_1^4 q_2^4 + L_2^4 q_1^4)-L_1^2 L_2^2 (c q_1^4-16 q_2^2 q_1^2+c q_2^4)]}{15 (c L_1^4 L_2^4)} \nn\\
&& ~~~
    +O(q_1^5,q_2^5) \Big]\ell^4
+\Big[ \frac{\pi^6 c (2 L_1^6-3 L_2^2 L_1^4+L_2^6)}{17010 L_1^6 L_2^6}
      -\frac{8 \pi^6 (L_1^2-L_2^2) [2 L_1^4 q_2^2-L_2^2(L_1^2+L_2^2) q_1^2]}{945 L_1^6 L_2^6} \nn\\
&& ~~~
      +\frac{4 \pi^6 [L_1^4 L_2^2 (q_1^3+2q_2^3)-2 L_1^6 q_2^3 -L_2^6 q_1^3] }{315 L_1^6 L_2^6}\\
&& ~~~
      -\frac{8 \pi^6 [ (c-16) ( 2 L_1^6 q_2^4 + L_2^6 q_1^4)  - L_1^4 L_2^2 (c q_1^4-32 q_1^2 q_2^2+2 (c-8) q_2^4) ]}{315 c L_1^6 L_2^6}
      +O(q_1^5,q_2^5) \Big]\ell^6 \nn\\
&& ~~~
      + \cdots \ell^8
      +O(\ell^{10}). \nn
\eea

\underline{Symmetrized relative entropies}
\bea \label{cl5}
&& S(\r_A(L_1,\phi_1),\r_A(L_2,\phi_2)) =
\frac{\pi^4 [L_1^2 (c-24 h_{\phi_2})-L_2^2 (c-24 h_{\phi_1})]^2\ell^4}{540 c L_1^4 L_2^4} \\
&& ~~~
+\frac{\pi^6 [L_1^2 (c-24 h_{\phi_2})-L_2^2 (c-24 h_{\phi_1})]^2 [L_1^2 (c-24 h_{\phi_2})+L_2^2 (c-24 h_{\phi_1})]\ell^6}{5670 c^2 L_1^6 L_2^6}
      + \cdots \ell^8
      +O(\ell^{10}), \nn
\eea
\bea \label{cl6}
&& S(\r_A(L_1,\phi),\r_A(L_2,q)) =
\Big[ \frac{\pi^4 [c L_1^2-(c-24 h_{\phi}) L_2^2]^2}{540 c L_1^4 L_2^4}
     +\frac{8 \pi^4 [(c-24h_{\phi}) L_2^2-c L_1^2] q^2}{45 c L_1^2 L_2^4} \nn\\
&& ~~~
     +\frac{4 \pi^4 [(c-24 h_{\phi}) L_2^2-c L_1^2] q^3}{15 c L_1^2 L_2^4}
     +\frac{8 \pi^4 [(c-24 h_{\phi}) L_2^2-(c-8) L_1^2] q^4}{15 c L_1^2 L_2^4}
     +O(q^5) \Big] \ell^4 \nn\\
&& ~~~
+\Big[ \frac{\pi^6 [c L_1^2-(c-24 h_{\phi}) L_2^2]^2 [c L_1^2+(c-24 h_{\phi}) L_2^2]}{5670c^2 L_1^6 L_2^6}  \\
&& ~~~
      -\frac{8 \pi^6 [c L_1^2-(c-24 h_{\phi}) L_2^2] [3 c L_1^2+(c-24 h_{\phi}) L_2^2] q^2}{945 c^2 L_1^4 L_2^6} \nn\\
&& ~~~
      -\frac{4 \pi^6 [c L_1^2-(c-24 h_{\phi}) L_2^2]
   [3 c L_1^2+(c-24 h_{\phi}) L_2^2] q^3}{315 c^2 L_1^4 L_2^6} \nn\\
&& ~~~
   -\frac{8 \pi^6 [3 c (c-16) L_1^4-2 (c-8) (c-24 h_{\phi}) L_1^2 L_2^2 -(c-24 h_{\phi})^2 L_2^4] q^4}{315 c^2 L_1^4 L_2^6}
   +O(q^5)\Big]\ell^6 \nn\\
&& ~~~
      + \cdots \ell^8
      +O(\ell^{10}), \nn
\eea
\bea \label{cl7}
&& S(\r_A(L_1,q_1),\r_A(L_2,q_2)) =
\Big[ \frac{c \pi^4 (L_1^2-L_2^2)^2}{540 L_1^4 L_2^4}
     -\frac{8 \pi^4 (L_1^2-L_2^2) (L_1^2 q_2^2-L_2^2 q_1^2)}{45 L_1^4L_2^4} \nn\\
&& ~~~
     -\frac{4 \pi^4 (L_1^2-L_2^2) (L_1^2 q_2^3-L_2^2q_1^3)}{15 L_1^4 L_2^4}
     +\frac{8 \pi^4 [8 (L_2^2 q_1^2-L_1^2 q_2^2)^2+c (L_1^2-L_2^2)(L_2^2 q_1^4-L_1^2 q_2^4)]}
           {15 c L_1^4 L_2^4}\\
&& ~~~
     +O(q_1^5,q_2^5) \Big]\ell^4
+\Big[ \frac{c \pi^6(L_1^2-L_2^2)^2 (L_1^2+L_2^2)}{5670 L_1^6 L_2^6}
      -\frac{8 \pi^6 (L_1^2-L_2^2)[3 q_2^2 L_1^4+L_2^2 (q_2^2-q_1^2) L_1^2-3 L_2^4 q_1^2]}{945 L_1^6 L_2^6}  \nn\\
&& ~~~
      -\frac{4 \pi^6(L_1^2-L_2^2) [3 q_2^3 L_1^4+L_2^2 (q_2^3-q_1^3) L_1^2-3 L_2^4 q_1^3]}{315 L_1^6 L_2^6}
      + \cdots  +O(q_1^5,q_2^5) \Big]\ell^6
      + \cdots \ell^8
      +O(\ell^{10}).\nn
\eea

\underline{The 2nd symmetrized relative entropies}
\bea \label{cl8}
&& S_2(\r_A(L_1,\phi_1),\r_A(L_2,\phi_2)) =
\frac{\pi^4 [L_1^2 (c-24 h_{\phi_2})-L_2^2 (c-24 h_{\phi_1})]^2\ell^4}{4608 c L_1^4L_2^4} \nn\\
&& ~~~
+\frac{\pi^6 [L_1^2 (c-24 h_{\phi_2})-L_2^2 (c-24 h_{\phi_1})] [L_1^4(11 c-240 h_{\phi_2})- L_2^4 (11c-240 h_{\phi_1})]\ell^6}
      {552960 c L_1^6 L_2^6} \nn\\
&& ~~~
  + \cdots \ell^8
  +O(\ell^{10}),
\eea
\bea\label{cl9}
&& S_2(\r_A(L_1,\phi),\r_A(L_2,q)) =
\Big[ \frac{\pi^4 [c L_1^2-(c-24 h_{\phi}) L_2^2]^2}{4608 c L_1^4 L_2^4}
     +\frac{\pi^4 [(c-24 h_{\phi}) L_2^2-c L_1^2] q^2}{48 c L_1^2 L_2^4}  \nn\\
&& ~~~
     +\frac{\pi^4 [(c-24 h_{\phi}) L_2^2-c L_1^2] q^3}{32 c L_1^2 L_2^4}
     +\frac{\pi^4 [(c-24 h_{\phi}) L_2^2-(c-8) L_1^2] q^4}{16 c L_1^2 L_2^4}+O(q^5) \Big] \ell^4  \nn\\
&& ~~~
+ \Big[ \frac{\pi^6 [c L_1^2-(c-24 h_{\phi}) L_2^2] [11 c L_1^4+(240 h_{\phi}-11 c)L_2^4]}
             {552960 c L_1^6 L_2^6}  \nn\\
&& ~~~
       +\frac{\pi^6 [3 c (10 c+33) L_1^4-10 (3 c+11) (c-24 h_{\phi}) L_2^2 L_1^2+(11 c-240h_{\phi}) L_2^4] q^2}
             {11520 c L_1^4 L_2^6}  \nn\\
&& ~~~
   +\frac{\pi^6 [c (80 c+319) L_1^4-10 (8 c+33) (c-24 h_{\phi}) L_2^2L_1^2+(11 c-240 h_{\phi}) L_2^4] q^3}
         {7680 c L_1^4 L_2^6}  \nn\\
&& ~~~
   +\frac{\pi^6 [11 (c (10 c+9)-160) L_1^4-10 (11 c+59) (c-24 h_{\phi}) L_2^2 L_1^2+(11 c-240 h_{\phi}) L_2^4] q^4}
         {3840 c L_1^4L_2^6} \nn\\
&& ~~~
   +O(q^5) \Big]\ell^6
   + \cdots \ell^8
   +O(\ell^{10}),
\eea
\bea\label{cl10}
&& S_2(\r_A(L_1,q_1),\r_A(L_2,q_2))=
\Big[ \frac{c \pi^4 (L_1^2-L_2^2)^2}{4608 L_1^4 L_2^4}
     -\frac{\pi^4 (L_1^2-L_2^2) (L_1^2 q_2^2-L_2^2q_1^2)}{48 L_1^4 L_2^4}\nn\\
&& ~~~
     -\frac{\pi^4 (L_1^2-L_2^2) (L_1^2 q_2^3-L_2^2 q_1^3)}{32 L_1^4 L_2^4}
     +\frac{\pi^4 [8 (L_2^2 q_1^2-L_1^2 q_2^2)^2+c (L_1^2-L_2^2)(L_2^2 q_1^4-L_1^2 q_2^4)]}{16 c L_1^4 L_2^4}\nn\\
&& ~~~+O(q_1^5,q_2^5) \Big]\ell^4
+\Big[ \frac{11\pi^6 c(L_1^2-L_2^2)^2 (L_1^2+L_2^2)}{552960 L_1^6 L_2^6}\nn\\
&& ~~~
      +\frac{\pi^6 (L_1^2-L_2^2) [3 (10 c+33) q_2^2 L_1^4+11 L_2^2 (q_1^2-q_2^2) L_1^2-3 (10 c+33) L_2^4 q_1^2]}{11520 L_1^6 L_2^6}\nn\\
&& ~~~
      +\frac{\pi^6 (L_1^2-L_2^2)[(80 c+319) q_2^3 L_1^4+11 L_2^2 (q_1^3-q_2^3) L_1^2-(80 c+319) L_2^4 q_1^3]}{7680 L_1^6 L_2^6}\nn\\
&& ~~~
   +\cdots + O(q_1^5,q_2^5) \Big]\ell^6
   + \cdots \ell^8
   +O(\ell^{10}).
\eea

\underline{The Jensen-Shannon divergence}
\bea \label{cl11}
&& JS(\r_A(L_1,\phi_1),\r_A(L_2,\phi_2)) =
\frac{\pi^4[L_1^2 (c-24 h_{\phi_2})-L_2^2 (c-24 h_{\phi_1})]^2\ell^4}{4320 c L_1^4 L_2^4} \nn\\
&& ~~~
+\frac{\pi^6 [L_1^2 (c-24 h_{\phi_2})-L_2^2 (c-24 h_{\phi_1})]^2 [L_1^2 (c-24 h_{\phi_2})+L_2^2 (c-24 h_{\phi_1})]\ell^6}{45360 c^2 L_1^6 L_2^6} \nn\\
&& ~~~
   + \cdots \ell^8
   +O(\ell^{10}),
\eea
\bea \label{cl12}
&& JS(\r_A(L_1,\phi),\r_A(L_2,q)) =
\Big[ \frac{\pi^4 [c L_1^2-(c-24 h_{\phi}) L_2^2]^2}{4320 c L_1^4 L_2^4}
     +\frac{\pi^4 [(c-24 h_{\phi}) L_2^2-c L_1^2] q^2}{45 c L_1^2 L_2^4}\nn\\
&& ~~~
     +\frac{\pi^4 [(c-24 h_{\phi}) L_2^2-c L_1^2]q^3}{30 c L_1^2 L_2^4}
     +\frac{\pi^4 [(c-24 h_{\phi}) L_2^2-(c-8) L_1^2] q^4}{15 c L_1^2 L_2^4}
     +O(q^5) \Big] \ell^4\nn\\
&& ~~~
+\Big[ \frac{\pi^6 [c L_1^2-(c-24 h_{\phi}) L_2^2]^2 [c L_1^2+(c-24 h_{\phi}) L_2^2]}{45360 c^2 L_1^6 L_2^6}\nn\\
&& ~~~
     -\frac{\pi^6 [c L_1^2-(c-24 h_{\phi}) L_2^2] [3 c L_1^2+(c-24 h_{\phi})
   L_2^2] q^2}{945c^2 L_1^4 L_2^6}\\
&& ~~~
     -\frac{\pi^6 [c L_1^2-(c-24 h_{\phi}) L_2^2] [3
   c L_1^2+(c-24 h_{\phi}) L_2^2] q^3}{630 c^2 L_1^4 L_2^6}\nn\\
&& ~~~
     -\frac{\pi^6 [3 (c-16) c L_1^4-2 (c-8) (c-24 h_{\phi}) L_1^2 L_2^2 -(c-24 h_{\phi})^2 L_2^4] q^4}{315 c^2 L_1^4 L_2^6}\nn\\
&& ~~~
     +O(q^5) \Big]\ell^6
     + \cdots \ell^8
     +O(\ell^{10}),\nn
\eea
\bea \label{cl13}
&& JS(\r_A(L_1,q_1),\r_A(L_2,q_2)) =
\Big[ \frac{c \pi^4 (L_1^2-L_2^2)^2}{4320 L_1^4 L_2^4}
     -\frac{\pi^4 (L_1^2-L_2^2) (L_1^2 q_2^2-L_2^2 q_1^2) x^2}{45 L_1^4 L_2^4} \nn\\
&& ~~~
     -\frac{\pi^4 (L_1^2-L_2^2) (L_1^2 q_2^3-L_2^2 q_1^3)}{30 L_1^4 L_2^4}
     +\frac{\pi^4 [8 (L_2^2 q_1^2-L_1^2 q_2^2)^2+c (L_1^2-L_2^2)(L_2^2 q_1^4-L_1^2 q_2^4)]}{15 c L_1^4 L_2^4}\\
&& ~~~
     +O(q_1^5,q_2^5) \Big]\ell^4
+\Big[ \frac{c \pi^6 (L_1^2-L_2^2)^2 (L_1^2+L_2^2)}{45360 L_1^6 L_2^6}
      -\frac{\pi^6 (L_1^2-L_2^2) (3 q_2^2   L_1^4+L_2^2 (q_2^2-q_1^2) L_1^2-3 L_2^4 q_1^2)}{945 L_1^6 L_2^6}\nn\\
&& ~~~
      -\frac{\pi^6 (L_1^2-L_2^2) [3 q_2^3 L_1^4+L_2^2 (q_2^3-q_1^3) L_1^2-3 L_2^4 q_1^3]}{630 L_1^6 L_2^6}
   +\cdots
   +O(q_1^5,q_2^5) \Big]\ell^6
   + \cdots \ell^8
   +O(\ell^{10}). \nn
\eea

\underline{The Schatten 2-norms}
\bea \label{cl14}
&& \|\r_A(L_1,\phi_1)-\r_A(L_2,\phi_2)\|_{2}^2 = \Big( \f{\ell}{\e} \Big)^{-\f{c}8} \Big\{
\frac{\pi^4 (c+2) \ell^4 [L_1^2 (c-24 h_{\phi_2})-L_2^2 (c-24 h_{\phi_1})]^2}
     {9216 c L_1^4 L_2^4} \nn\\
&& ~~~
+\frac{\pi^6 (c+4) \ell^6}{4423680 c L_1^6 L_2^6}
[L_1^2 (c-24 h_{\phi_2})-L_2^2 (c-24 h_{\phi_1})]
\{      L_1^4 [c (5 c+22)-240 h_{\phi_2} (c-12 h_{\phi_2}+2)]\nn\\
&& ~~~
      - L_2^4 [c (5c+22)-240 h_{\phi_1} (c-12 h_{\phi_1}+2)]\}
   + \cdots \ell^8
+O(\ell^{10})
\Big\},
\eea
\bea \label{cl15}
&& \|\r_A(L_1,\phi)-\r_A(L_2,q)\|_{2}^2 = \Big( \f{\ell}{\e} \Big)^{-\f{c}8} \bigg\{
\pi^4(c+2)\Big\{ \frac{ [c L_1^2-(c-24 h_{\phi}) L_2^2]^2}{9216 c L_1^4 L_2^4} \nn\\
&& ~~~
     +\frac{[(c-24 h_{\phi}) L_2^2-c L_1^2 ]q^2}{96 c L_1^2 L_2^4}
     +\frac{[(c-24 h_{\phi}) L_2^2-c L_1^2] q^3}{64 c L_1^2 L_2^4}
     +\frac{[(c-24 h_{\phi}) L_2^2-(c-8) L_1^2]q^4}{32 c L_1^2L_2^4}\nn\\
&& ~~~ +O(q^5)
\Big\}\ell^4
+\pi^6(c+4)\Big\{ \frac{[c L_1^2-(c-24 h_{\phi}) L_2^2]}{4423680 c L_1^6 L_2^6}\{ c (5 c+22)L_1^4 \nn\\
&& ~~~+[240 (c-12 h_{\phi}+2) h_{\phi}-c (5 c+22)] L_2^4 \}
    +\frac{ q^2}{92160 c L_1^4 L_2^6}\{ 9 c (5 c+22) L_1^4\\
&& ~~~-10 (5 c+22) (c-24 h_{\phi})L_1^2 L_2^2+[c (5 c+22)-240 (c-12 h_{\phi}+2) h_{\phi}]L_2^4 \}\nn\\
&& ~~~
     +\frac{q^3}{61440 c L_1^4 L_2^6}
     \{ 29 c (5 c+22) L_1^4-30 (5 c+22) (c-24 h_{\phi}) L_2^2
   L_1^2\nn\\
&& ~~~+[c (5 c+22)-240 (c-12 h_{\phi}+2) h_{\phi}] L_2^4 \}
   +\cdots q^4 + O(q^5)
\Big\}\ell^6
+\cdots \ell^8+O(\ell^{10})\nn
\bigg\},
\eea
\bea \label{cl16}
&& \|\r_A(L_1,q_1)-\r_A(L_2,q_2)\|_{2}^2 =\Big( \f{\ell}{\e} \Big)^{-\f{c}8} \Big\{
\pi^4(c+2)\Big[ \frac{c (L_1^2-L_2^2)^2}{9216 L_1^4 L_2^4}\nn\\
&& ~~~
     -\frac{(L_1^2-L_2^2) (L_1^2 q_2^2-L_2^2 q_1^2)}
           {96L_1^4 L_2^4}
     -\frac{(L_1^2-L_2^2) (L_1^2 q_2^3-L_2^2 q_1^3)}
           {64 L_1^4 L_2^4}\\
&& ~~~
     -\frac{ (c-8)( L_1^4 q_2^4 + L_2^4 q_1^4)-L_1^2 L_2^2 (c q_1^4-16 q_2^2 q_1^2+c q_2^4)}
           {32 c L_1^4 L_2^4}
   +O(q_1^5,q_2^5) \Big]\ell^4 \nn\\
&& ~~~
+ \pi^6 c (c+4) (5 c+22)\Big[ \frac{L_1^4-L_2^4 }{4423680 L_1^6 L_2^6}
      +\frac{ 9 L_1^4 q_2^2 +L_1^2 L_2^2 (q_1^2-q_2^2) -9 L_2^4 q_1^2}{92160 L_1^6 L_2^6}\nn\\
&& ~~~
      +\frac{29 L_1^4 q_2^3 +L_1^2 L_2^2 (q_1^3-q_2^3) -29
   L_2^4 q_1^3}{61440 L_1^6 L_2^6} + \cdots +O(q_1^5,q_2^5)  \Big]\ell^6
   +\cdots\ell^8
   +O(\ell^{10})
\Big\}. \nn
\eea

\underline{The Schatten 4-norms}
\bea  \label{cl17}
&& \|\r_A(L_1,\phi_1)-\r_A(L_2,\phi_2)\|_{4}^4 = \Big( \f{\ell}{\e} \Big)^{-\f{5c}{16}} \Big[
\frac{\pi^8 (c+2) [25 c (25 c-14)+192]\ell^8}{21743271936
   c^3 L_1^8 L_2^8}[L_1^2 (c-24 h_{\phi_2}) \nn\\ && ~~~ -L_2^2 (c-24 h_{\phi_1})]^4
+O(\ell^{10})
\Big],
\eea
\bea \label{cl18}
&& \|\r_A(L_1,\phi)-\r_A(L_2,q)\|_{4}^4 = \Big( \f{\ell}{\e} \Big)^{-\f{5c}{16}} \Big\{
\pi^8 (c+2) [25 c (25 c-14)+192] \Big[
\frac{ [c L_1^2-L_2^2 (c-24 h_{\phi})]^4}{21743271936 c^3 L_1^8 L_2^8} \nn\\
&& ~~~
-\frac{  [c L_1^2-L_2^2 (c-24 h_{\phi})]^3 q^2}{113246208 c^3 L_1^6 L_2^8}
-\frac{  [c L_1^2-L_2^2 (c-24 h_{\phi})]^3q^3}{75497472
   c^3 L_1^6 L_2^8}\\
&& ~~~
-\frac{  [(c-24) L_1^2-L_2^2 (c-24 h_{\phi})]
   [c L_1^2-L_2^2 (c-24 h_{\phi})]^2q^4}{37748736 c^3 L_1^6 L_2^8}
+O(q^5)
\Big]\ell^8 +O(\ell^{10})\nn
\Big\},
\eea
\bea \label{cl19}
&& \|\r_A(L_1,q_1)-\r_A(L_2,q_2)\|_{4}^4 = \Big( \f{\ell}{\e} \Big)^{-\f{5c}{16}} \Big\{
\pi^8(c+2) [25 c (25 c-14)+192] \Big[
\frac{ c (L_1^2-L_2^2)^4}{21743271936 L_1^8 L_2^8} \nn\\
&& ~~~
-\frac{(L_1^2-L_2^2)^3 (L_1^2 q_2^2-L_2^2 q_1^2)}{113246208 L_1^8 L_2^8}
-\frac{(L_1^2-L_2^2)^3 (L_1^2 q_2^3-L_2^2 q_1^3)}{75497472 L_1^8
   L_2^8}\nn\\
&& ~~~
-\frac{(L_1^2-L_2^2)^2 [(c-24) ( L_1^4
   q_2^4 + L_2^4 q_1^4 )-L_1^2 L_2^2 (c q_1^4+c q_2^4-48 q_2^2 q_1^2)]}{37748736 c L_1^8 L_2^8}
\nn\\
&& ~~~
+O(q_1^5,q_2^5)
\Big]\ell^8 +O(\ell^{10})
\Big\}.
\eea

\providecommand{\href}[2]{#2}\begingroup\raggedright\endgroup


\end{document}